\documentclass[lettersize,journal]{IEEEtran}

\usepackage{amsmath,amsfonts}
\usepackage{algorithmic}
\usepackage{algorithm}
\usepackage{array}
\usepackage[caption=false, font=scriptsize, labelfont=rm, textfont=rm]{subfig}
\usepackage{textcomp}
\usepackage{stfloats}
\usepackage{url}
\usepackage{verbatim}
\usepackage{graphicx}
\usepackage{cite}
\usepackage{amsmath,amssymb,amsfonts}
\usepackage{graphicx}
\usepackage{textcomp}
\usepackage{xcolor}
\usepackage{booktabs}
\usepackage{multirow}
\usepackage{booktabs}  % 必须：用于画三线表
\usepackage{multirow}  % 用于合并单元格
\usepackage{xcolor}    % 用于高亮关键词
\usepackage{tabularx}  % 用于自动换行
\usepackage{amssymb}
\usepackage{hyperref}
\usepackage{url}

\newcommand{\softplus}{\operatorname{softplus}}

\begin{document}

\title{LogICL: Demonstration Retrieval to Bridge the Semantic Gap in Cross-Domain Log Anomaly Detection}

\author{Jingwei~Ye,
        Zhi~Wang,
        % Zhi~Wang,~\IEEEmembership{Member,~IEEE},
        Chenbin~Su,
        Jieshuai~Yang,
        Jiayi~Ding,
        Chunbo~Liu,
        and Ge~Chu,~\IEEEmembership{Member,~IEEE}
        % <-this % stops a space
\thanks{ (Corresponding author: Zhi Wang.)}%
        % 第二个 \thanks：写 Ye Jingwei 和 Zhi Wang 的单位及邮箱
\thanks{Jingwei Ye, Zhi Wang, Chenbin Su, Jieshuai Yang are with the College of Cryptology and Cyber Science, Nankai University, Tianjin 300350, China (e-mail: jwye@mail.nankai.edu.cn; zwang@nankai.edu.cn; champion.su@mail.nankai.edu.cn;  yjs@mail.nankai.edu.cn).}%
\thanks{Jiayi Ding is with the College of Safety Science and Engineering, Civil Aviation University of China, Tianjin 300300, China (e-mail: 221240005@cauc.edu.cn)}
\thanks{Chunbo Liu is with the Information Security Evaluation Center, Civil Aviation University of China, Tianjin 300300, China (e-mail: luchamber@163.com)}
\thanks{Ge Chu is with the Research Department, Runjian Co., Ltd., Nanning 530200, China (e-mail: gechu@runjian.com)}

        % 第三个 \thanks：写 Wang Wu 的单位及邮箱 (如果单位不同)
}

% Journal header and publication identifier are intentionally omitted in the
% author manuscript. They should be restored only from publisher-supplied text.

\maketitle

\begin{abstract}

Effective log anomaly detection is critical to sustaining reliability in large-scale IT infrastructures. Transformer-based language models have advanced the field but require substantial computational resources and labeled data, exacerbating the cold-start problem in target domains where logs are scarce due to system evolution.
Cross-domain methods alleviate data scarcity by leveraging source domain logs. However, their reliance on surface lexical similarity limits cross-domain generalization, as they fail to capture latent semantic equivalence amid structural and terminological divergences.

To address these challenges, we propose LogICL, a novel demonstration-retrieval
 framework for cross-domain log anomaly detection. During training, LogICL 
 constructs a sparse delta matrix by measuring the signed change in a frozen 
 LLM's prediction error between zero-shot and one-shot inference for selected 
 query--demonstration pairs. This outcome-level feedback supervises a 
 lightweight encoder through a demonstration-utility loss, while maximum mean 
 discrepancy and supervised contrastive losses promote domain alignment and 
 anomaly discrimination, respectively. At inference, the trained encoder first 
 retrieves similarity-based anchors and then expands them with candidates having
 the highest scores aggregated over precomputed positive delta relations. This procedure retrieves high-utility demonstrations 
 without candidate-wise LLM evaluation and supplies them to a frozen LLM for 
 in-context anomaly prediction. 
 
 Experiments on few-shot and target-domain zero-shot transfer benchmarks demonstrate 
 strong performance across heterogeneous log systems. In the few-shot setting, LogICL
 outperforms both Random-ICL and $k$NN-ICL in three of the four transfers under the fixed
 split, while the benefit remains setting-dependent. Further analyses indicate that 
 LogICL can identify demonstrations with positive predictive utility despite 
 low Surface Lexical Similarity, thereby helping bridge the semantic gap across logging 
 systems.

\end{abstract}

\begin{IEEEkeywords}
Log anomaly detection, Cross-domain, In-context learning, Latent semantic equivalence, Large language model.
\end{IEEEkeywords}

\section{Introduction}
Modern large-scale software systems—including distributed platforms, cloud services, and enterprise infrastructures—generate massive volumes of logs that record system operations, component interactions, and runtime states. As a primary source of observability, these logs are critical for maintaining system reliability and play a central role in diagnosing performance degradation, spotting service failures, and detecting anomalies. Therefore, effective log analysis is essential for ensuring the robustness and trustworthiness of modern computing systems.

Log-based anomaly detection has therefore become an indispensable technique for maintaining stable and dependable software operations, enabling timely identification and diagnosis of abnormal behaviors before they escalate into severe system disruptions~\cite{chalapathy2019deep,he2021survey,he2016experience}. 

In recent years, Transformer-based language models have established a new paradigm in log anomaly detection, broadly categorized into Language Models (LMs) and Large Language Models (LLMs). LM-based approaches (e.g., LogBERT~\cite{guo2021logbert}, NeuralLog~\cite{le2021log}, LogOW~\cite{ye2025logow}) leverage masked language modeling and contextual embeddings to capture stable token-level patterns, enabling robust anomaly scoring with modest supervision. More recently, LLM-based methods (e.g., LogGPT~\cite{han2023loggpt}, LogLLM~\cite{guan2024logllm}) have demonstrated remarkable capabilities in reasoning and few-shot generalization, utilizing high-capacity encoders and extensive pre-training to model complex, long-range log dependencies across diverse formats.

% However, the efficacy of these Transformer-based architectures comes at a significant cost. Both LMs and LLMs demand substantial computational resources and large-scale, high-quality datasets for pre-training or fine-tuning. While LLMs exhibit superior few-shot performance, their massive parameter scale introduces high latency and deployment costs. Even relatively smaller LMs require considerable training overhead compared to lightweight statistical methods, limiting their feasibility in resource-constrained or real-time operational environments. 

% However, these Transformer-based approaches achieve high accuracy and generalization at a substantial cost: both LMs and LLMs rely heavily on large-scale, high-quality labeled datasets for effective pre-training or fine-tuning. In the absence of sufficient labeled data—which is common in new or cross-domain deployments—their performance degrades sharply. Moreover, their substantial parameter counts and training requirements impose significantly higher computational overhead than lightweight statistical or deep-log parsers, rendering them impractical when labeled logs are scarce or computational resources are limited.
% This challenge is further exacerbated by the rapid evolution of modern software systems~\cite{torrey2010transfer, ye2025logow}. The scarcity of labeled logs in target systems renders the supervised training of data-intensive Transformer-based log anomaly detection models infeasible~\cite{ye2025logow}. 

However, the superior performance of Transformer-based architectures comes at a significant cost. These models are data-intensive and computationally demanding, necessitating extensive labeled datasets for effective pre-training or fine-tuning. This dependency creates a fundamental conflict with the rapid evolution of modern software systems, where continuous updates lead to a scarcity of labeled logs in newly deployed environments~\cite{torrey2010transfer, ye2025logow}. Consequently, the substantial parameter scale and training overhead render these models practically infeasible in data-scarce target domains.

To address data scarcity, cross-system methods attempt to transfer knowledge from high-resource source domains to target domains. LogTransfer~\cite{chen2020logtransfer} uses shared networks to enable cross-system anomaly detection; LogTAD~\cite{han2021unsupervised} employs adversarial domain adaptation techniques for cross-domain training; MetaLog~\cite{zhang2024metalog} enables cross-system anomaly detection by leveraging meta-learning on globally consistent semantic embeddings.

Although existing methods have achieved preliminary success in mitigating data scarcity, they continue to face significant challenges in cross-domain scenarios. These challenges can be summarized in two key areas:

\paragraph{Surface Similarity vs. Latent Equivalence}
Existing cross-domain log anomaly detection methods predominantly leverage static pre-trained word embeddings (e.g., GloVe~\cite{pennington2014glove}) to obtain semantic vector representations. While effective for logs with high lexical overlap, these approaches capture only \textit{surface lexical similarity}, failing to model the \textit{latent semantic equivalence} required for heterogeneous systems.

\begin{figure}[htbp]
\centerline{\includegraphics[scale=0.7]{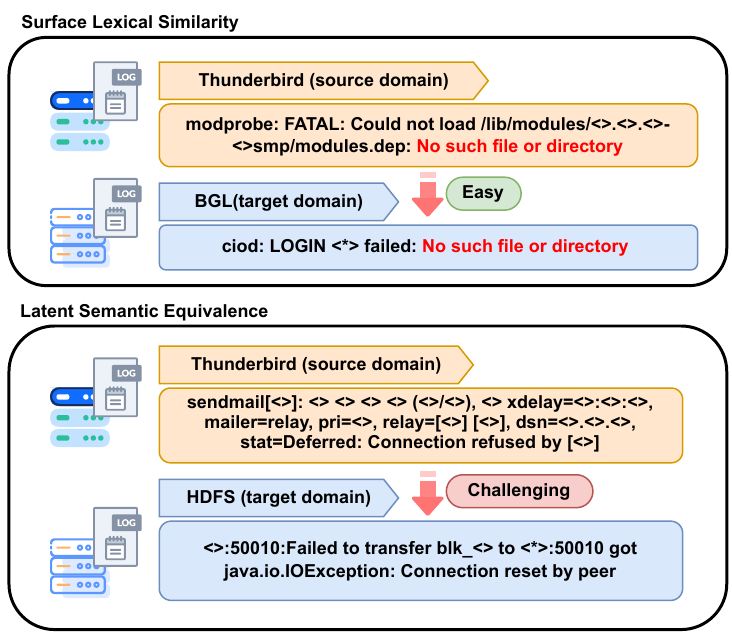}}
\caption{Surface Lexical Similarity \& Latent Semantic Equivalence.}
\label{fig:surface_vs_latent}
\end{figure}

The upper part of Figure~\ref{fig:surface_vs_latent} demonstrates a scenario where static embeddings succeed due to substantial consistency. 
From the perspective of \textbf{syntactic structure}, both the Thunderbird (source domain) and BGL (target domain) logs adhere to a unified \textit{Component-State-Reason} template. The source begins with a kernel component (``modprobe'') followed by a severity level (``FATAL''), while the target begins with a component header (``ciod'') followed by an action state (``LOGIN ... failed''), and both conclude with the identical error reason (``No such file or directory'').
In terms of \textbf{lexical vocabulary}, the two domains exhibit high overlap, sharing nearly identical error descriptors (e.g., the exact POSIX error ``No such file or directory''). This structural and lexical alignment allows traditional methods to easily extract correlated features and facilitate positive transfer.

In contrast, the lower part of Figure~\ref{fig:surface_vs_latent} highlights the limitation of existing methods in identifying a related \textit{Network Transmission Failure} across heterogeneous systems.
From the perspective of \textbf{syntactic structure}, the domains show fundamental divergence. The Thunderbird (source domain) log employs a \textit{declarative key-value} format, listing component attributes (``sendmail'' and ``mailer=relay'') and a status indicator (``stat=Deferred''). Conversely, the HDFS (target domain) log adopts a \textit{procedural narrative} typical of Java exceptions, describing a failed block-transfer operation (``Failed to transfer blk\_...'') that culminates in ``java.io.IOException: Connection reset by peer.''
In terms of \textbf{lexical vocabulary}, the two logs use disjoint expressions (``relay/refused'' vs. ``transfer/reset'') to describe related network-transmission failures. This significant discrepancy in both structure and vocabulary makes the cross-domain relation difficult to capture using surface lexical similarity alone.

\paragraph{High Computational and Data Constraints of Language Models}
While Transformer-based models offer superior performance, their deployment is hindered by excessive resource demands. This constraint applies not only to massive LLMs but also to BERT-based LMs, which still require significant computational power for fine-tuning and inference compared to traditional methods. The dependence on high-quality labeled datasets further complicates their application in real-world environments where data are noisy or scarce. Even with efficient tuning strategies (e.g., LoRA~\cite{hu2022lora}), the fundamental trade-off between model capacity and resource efficiency remains a persistent hurdle for scalable log analysis.

% To resolve the limitations outlined above, we propose LogICL, a framework that leverages in-context learning (ICL) for improved cross-domain generalization. Specifically, LogICL addresses two critical challenges: (1) inadequate modeling of semantic equivalence across logs with varying structures and terminology, and (2) the high computational and data demands of large language models (LLMs), which typically require extensive fine-tuning and labeled datasets. In summary, this paper makes the following contributions:

In summary, to address the aforementioned limitations, we propose LogICL, a novel framework that harnesses ICL to enhance cross-domain generalization and detection accuracy in few-shot log anomaly detection settings with scarce target-domain supervision.
The main contributions of this paper are summarized as follows:

\begin{itemize}
    \item We formulate cross-domain log transfer as a demonstration-retrieval
    problem for in-context learning with a frozen LLM. LogICL uses the signed 
    change in prediction error from zero-shot to one-shot inference as an 
    offline, model-specific supervision signal, rather than relying solely on 
    surface lexical similarity.

    \item We train a lightweight log encoder using a multi-objective function 
    that combines demonstration-utility supervision, domain alignment, and 
    class discrimination. The resulting retrieval space reflects both 
    cross-domain compatibility and the observed predictive utility of 
    query--demonstration pairs without updating the LLM parameters.

    \item We develop a two-stage retrieval strategy that combines 
     similarity-based anchors with precomputed delta-based expansion, avoiding 
     candidate-wise LLM evaluation at test time. Under the stated fixed data 
     splits, the few-shot experiments compare this strategy with random and 
     similarity-only ICL retrieval, while the target-domain zero-shot transfer 
     experiments evaluate transfer without pre-evaluation access to target-domain data.
\end{itemize}

\section{Preliminaries and Related Work}

\subsection{Preliminaries}

Transfer learning (TL) improves generalization in a target domain by reusing knowledge acquired from related source domains~\cite{weiss2016survey,zhuang2020comprehensive,DBLP:journals/tkde/PanY10}. Existing approaches generally transfer instances, feature representations, or model parameters. Feature-based transfer is particularly relevant to cross-domain log analysis because log formats and vocabularies often differ substantially across systems. LogICL follows this setting by learning a shared log representation that reduces inter-domain discrepancy while preserving the information required to distinguish normal and anomalous sequences.

In-context learning (ICL) enables a pretrained language model to perform a task by conditioning on a small set of input--output demonstrations without updating its parameters~\cite{dong2022survey,pan2023context,brown2020language}. As illustrated in Fig.~2, labeled log sequences are concatenated with a query to guide a frozen LLM in predicting whether the query is normal or anomalous. ICL performance, however, is highly sensitive to demonstration selection: relevant examples can improve prediction, whereas unsuitable ones may introduce interference~\cite{liu2022makes,peng2024revisiting,qin2024context}. Effective selection should therefore consider not only textual similarity but also the utility of a candidate to the inference model.

\subsection{Related Work}

Traditional log anomaly detection commonly comprises log collection and preprocessing, parsing and grouping, representation learning, and anomaly scoring~\cite{he2021survey,he2016experience}. Recent LLM-based studies extend different parts of this pipeline. LLMParser~\cite{ma2024llmparser}, LogPrompt~\cite{liu2024logprompt}, and LogRules~\cite{huang2025logrules} apply few-shot tuning, prompting, and rule induction, respectively, to log parsing or broader log analysis. LogRAG~\cite{zhang2024lograg} and EagerLog~\cite{duan2025eagerlog} incorporate retrieval-augmented generation into anomaly detection, with the latter further employing active learning. Other methods, including LogGPT~\cite{han2023loggpt}, LogLLaMA~\cite{yang2025logllama}, LogLLM~\cite{guan2024logllm}, and SuperLog~\cite{ji2025adapting}, adapt trainable model components to log-specific tasks. In particular, LogLLaMA fine-tunes LLaMA2 on normal log-key sequences for next-event prediction and further applies reinforcement learning for anomaly detection. WebNorm~\cite{liao2024detecting} addresses web tamper attacks by using LLM-assisted program and log analysis to learn executable consistency rules from web-application logs; violations of these rules both detect and explain tamper-attack anomalies. These studies mainly address log parsing, model adaptation, detection and explanation of web-tamper anomalies, or retrieval of supporting log evidence rather than cross-domain ICL demonstration selection.

Early retrieval-based ICL methods select examples that are semantically similar to the test query~\cite{liu2022makes}. Skill-KNN~\cite{an2023skill} reduces sensitivity to task-irrelevant surface features by retrieving over LLM-generated skill descriptions. More model-aware methods include TopK+ConE~\cite{peng2024revisiting}, which reranks a TopK candidate set using conditional entropy estimated by the inference model, and IDS~\cite{qin2024context}, which iteratively updates demonstrations using reasoning paths generated by zero-shot chain-of-thought prompting. These studies indicate that demonstration quality depends on both query relevance and compatibility with the inference model; however, model-aware reranking or iterative selection generally requires additional inference-model calls during selection.

Another line of research learns demonstration retrievers from offline LM feedback. EPR~\cite{rubin2022learning} trains a dense retriever from candidate preferences derived from the likelihood of the ground-truth output, while UDR~\cite{li2023unified} converts feedback collected across heterogeneous tasks into a unified listwise ranking objective. UPRISE~\cite{cheng2023uprise} trains a transferable prompt retriever using feedback from a frozen auxiliary LM, whereas LLM-R~\cite{wang2024learning} combines LLM-based candidate ranking, reward modeling, and knowledge distillation. GenICL~\cite{zhang2025learning} further learns generative demonstration preferences directly from LLM feedback. LogICL is most closely related to this offline-feedback family but specializes it for cross-domain log anomaly detection: the signed reduction in prediction error from zero-shot to one-shot inference supervises a domain-aligned and class-discriminative log encoder. At test time, the trained encoder and precomputed delta matrix support demonstration construction without candidate-wise LLM evaluation.

\section{Model Design}

\begin{figure*}[htbp]
\centering
\includegraphics[width=1\textwidth]{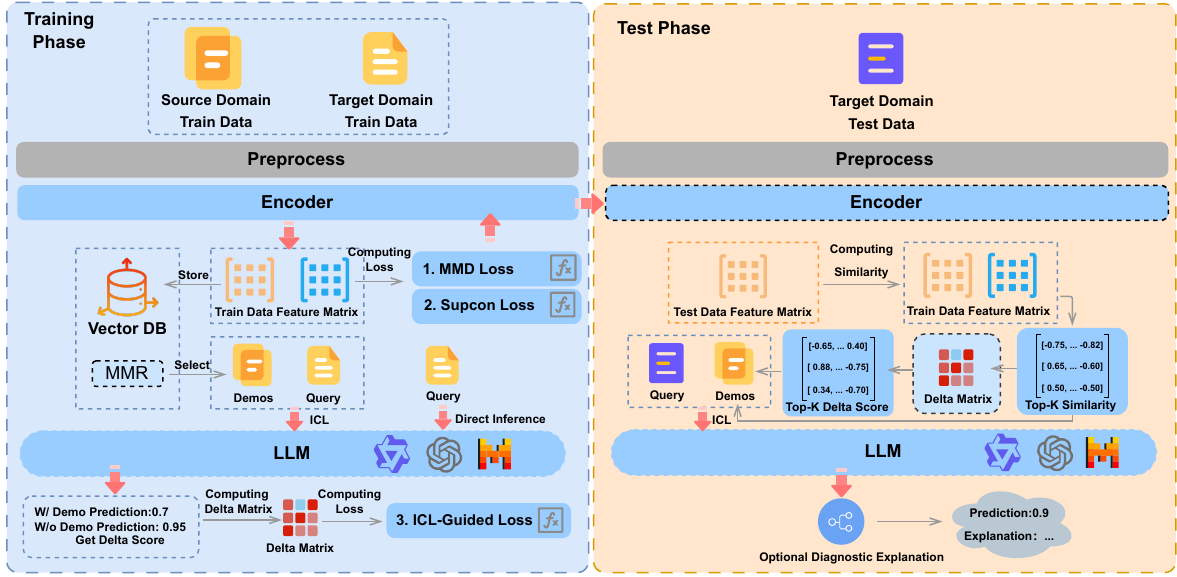}
\caption{Overall framework of LogICL.}
\label{fig:logicl_framework}
\end{figure*}

\subsection{Overview}
As illustrated in Fig.~\ref{fig:logicl_framework}, LogICL is a
demonstration-retrieval framework for cross-domain log anomaly detection with
a frozen LLM. The framework comprises a training phase and a test phase.
During training, Phase~1 first adapts the Sentence-BERT encoder using maximum
mean discrepancy and supervised contrastive losses to promote domain alignment
and anomaly discrimination. After Phase~1 adaptation, MMR identifies a sparse
set of candidate query--demonstration pairs. For each observed pair, LogICL
measures the signed reduction in the frozen LLM's prediction error when the
candidate is added to a zero-shot prompt, producing a fixed sparse delta
matrix. Phase~2 then uses this outcome-level utility signal to further refine
the encoder through the demonstration-utility loss $\mathcal{L}_{\Delta}$.

At test time, the optimized encoder retrieves similarity-based anchors and uses their precomputed positive delta relations to expand the demonstration set. The retrieved demonstrations are then supplied to the frozen LLM for in-context prediction. 

\subsection{Pre-processing}

Traditional log analysis employs parsers (e.g., Drain~\cite{he2017drain}) to split raw log messages into constant \textit{event templates} and dynamic \textit{parameters}. However, this parsing process introduces inherent limitations. First, parsing algorithms may misidentify variable parameters as template tokens. This error generates fragmented templates for logs belonging to the same event category, artificially introducing ``unseen formats'' that inject noise into downstream detection tasks~\cite{le2021log, jiang2024large, jiang2024lilac, le2023log}. Second, discarding dynamic parameters results in the loss of semantic information that is valuable for detection accuracy. Recent studies indicate that Transformer-based models can achieve robust performance directly on raw log text without explicit parsing~\cite{guan2024logllm, le2021log}.

Consequently, we adopt a \textbf{parser-free strategy}. We utilize regular expressions for basic preprocessing, which avoids parsing-induced noise while preserving raw parameter details (e.g., IP addresses and \textit{block\_IDs}). Crucially, retaining these parameters enables the downstream LLM to leverage the full semantic context during inference, thereby enhancing detection accuracy and providing interpretable diagnostics that help operations experts narrow the scope of anomaly localization.

% While traditional methods utilize parsers (e.g., Drain~\cite{he2017drain}) to abstract dynamic parameters into constant tokens, this approach often discards semantic context critical for anomaly localization and generalizes poorly to unseen formats~\cite{chen2021neurallog}. Conversely, LLMs leverage such parameter information for deeper contextual reasoning. Corroborating findings that lightweight preprocessing yields comparable performance to complex parsing for LLM-based tasks~\cite{guan2024logllm}, we adopt a parser-free strategy. We employ minimal regular expressions for noise filtering and normalization, preserving raw semantic content to maximize the reasoning and interpretability of the downstream LLM.

\subsection{Log Sequence Representation Learning}

Following preprocessing, raw log messages are grouped into sequences to capture system execution flows. This is achieved via two strategies: session-based partitioning, which aggregates messages by identifiers (e.g., \textit{block\_IDs} in HDFS), and window-based partitioning (fixed or sliding) for continuous streams. This structuring is critical, as it enables the model to detect anomalies based on sequential dependencies and behavioral patterns, rather than analyzing isolated log entries.

Let a log sequence be denoted as $s=\{ m_1, m_2, \dots, m_L \}$, where $s$ is a sequence of log messages $m_i$, and $L$ is the sequence length. Let $\mathcal{S}$ denote the space of log sequences. A training dataset of sequences is defined as $\mathcal{D}_{\text{train}} = \{ s_1, s_2, \dots, s_N \}$, where $s_i \in \mathcal{S}$ and $N$ is the number of sequences in the training dataset. Each log sequence $s_i$ is encoded into a fixed-dimensional embedding vector via the encoder $f_{\theta}$:
\begin{equation}
\begin{aligned}
v_i &= f_\theta(s_i) \in \mathbb{R}^d, \quad i = 1, \dots, N,
\end{aligned}
\label{eq:sentencebert}
\end{equation}
% \[
%   v_i = f_\theta(S_i) \in \mathbb{R}^d, \quad i = 1, \dots, N,
% \] 
where $v_i$ denotes the embedding vector of sequence $s_i$ and $d$ is the embedding dimension (i.e., the number of features representing each sequence).

Collectively, the set of embeddings for the training dataset is
\begin{equation}
\begin{aligned}
\mathcal{E} = \{ v_1, v_2, \dots, v_N \} \subset \mathbb{R}^d,
\end{aligned}
\label{eq:feature vector}
\end{equation}
where $\mathcal{E}$ represents the embedding space of all training sequences.

In cross-domain log anomaly detection, robust representation learning hinges on extracting domain-invariant features. Ideally, the encoder should suppress irrelevant system-specific variations via domain alignment to facilitate transfer. However, an inherent tension exists: aggressive alignment can precipitate negative transfer by erasing task-critical semantics. Since anomalies in heterogeneous systems often manifest through distinct keywords (e.g., \textit{block lost} in HDFS vs. \textit{file system full} in BGL), indiscriminate alignment risks treating these discriminative features as domain noise. Therefore, an effective framework must navigate this trade-off, bridging the distributional gap while rigorously preserving the semantic discriminability of system-specific anomalies.

\subsection{Training Phase}

LogICL adopts sequential two-stage encoder training while keeping the LLM frozen. Phase~1 adapts Sentence-BERT using domain-alignment and supervised contrastive objectives. The adapted encoder then selects candidate query--demonstration pairs for one-time delta-matrix construction, after which Phase~2 refines the encoder using the fixed LLM-derived utility signal. Let $\mathcal{S}_{\mathrm{train}}=\{(s_i,y_i)\}_{i=1}^{N}$ denote the labeled training sequences, where $y_i\in\{0,1\}$ is the normal/anomalous label. It contains source and limited target samples in few-shot transfer, and primary and auxiliary source samples in target-domain zero-shot transfer.

The overall stage-conditioned objective is

\begin{equation}
\label{eq:overall_training_loss}
\mathcal{L}_{\mathrm{Train}}^{(r)}
=
\begin{cases}
\lambda_{\mathrm{MMD}}\mathcal{L}_{\mathrm{MMD}}
+\lambda_{\mathrm{SupCon}}\mathcal{L}_{\mathrm{SupCon}}, & r=1,\\[3pt]
\lambda_{\Delta}\mathcal{L}_{\Delta}, & r=2,
\end{cases}
\end{equation}

where $r\in\{1,2\}$ indexes the training stage, and $\lambda_{\mathrm{MMD}}=0.1$, $\lambda_{\mathrm{SupCon}}=1.0$, and $\lambda_{\Delta}=1.0$. Phase~2 is initialized with the Phase~1 parameters, and its delta matrix and observed pair set remain fixed. Equation~(\ref{eq:overall_training_loss}) defines the objective used at each stage in this training order.

\subsubsection{Phase 1: Encoder Training with Domain Alignment and Supervised Contrastive Learning}

We initialize $f_{\boldsymbol{\theta}^{(0)}}$ from the pretrained Sentence-BERT model and optimize the first branch of Eq.~(\ref{eq:overall_training_loss}) for 50 epochs. The resulting encoder $f_{\boldsymbol{\theta}^{(1)}}$, rather than the off-the-shelf model, is used for the subsequent candidate selection.

\paragraph{Domain Alignment Loss}
Let $\mathcal{B}_{s}$ and $\mathcal{B}_{t}$ be mini-batches of sizes $N_s$ and $N_t$ from the two training domains. Maximum Mean Discrepancy (MMD) aligns their embedding distributions~\cite{gretton2012kernel,sriperumbudur2010hilbert}:

\begin{equation}
\label{eq:mmd_loss}
\mathcal{L}_{\mathrm{MMD}}
=
\left\|
\frac{1}{N_s}\sum_{i=1}^{N_s}
\phi\!\left(f_{\boldsymbol{\theta}}(s_i^{s})\right)
-
\frac{1}{N_t}\sum_{j=1}^{N_t}
\phi\!\left(f_{\boldsymbol{\theta}}(s_j^{t})\right)
\right\|_{\mathcal{H}},
\end{equation}

where $\phi(\cdot)$ is the feature map of a Gaussian kernel with bandwidth $\sigma=1.0$, and $\mathcal{H}$ is the corresponding reproducing kernel Hilbert space. Minimizing $\mathcal{L}_{\mathrm{MMD}}$ suppresses domain-specific distributional variation.

\paragraph{Supervised Contrastive Loss}
To preserve class structure during alignment, Supervised Contrastive Loss (SupCon) pulls same-label samples together and separates different-label samples~\cite{khosla2020supervised}. For a mini-batch $\mathcal{B}$, let $A(i)=\mathcal{B}\setminus\{i\}$, $P(i)=\{p\in A(i):y_p=y_i\}$, and $I=\{i\in\mathcal{B}:|P(i)|>0\}$. We define

\begin{equation}
\label{eq:supcon_loss}
\begin{aligned}
\mathcal{L}_{\mathrm{SupCon}}
= {}& -\frac{1}{|I|}\sum_{i\in I}\frac{1}{|P(i)|}
\sum_{p\in P(i)} \\
& \times\;\log\frac{
\exp\!\left(\operatorname{sim}(v_i,v_p)/\tau_c\right)
}{
\sum_{a\in A(i)}
\exp\!\left(\operatorname{sim}(v_i,v_a)/\tau_c\right)
},
\end{aligned}
\end{equation}

where $v_i=f_{\boldsymbol{\theta}}(s_i)$, $\operatorname{sim}(\cdot,\cdot)$ is cosine similarity, and $\tau_c=0.1$. Together, MMD and SupCon reduce dependence on domain-specific lexical patterns while preserving normal/anomalous discrimination. They provide a stronger initial retrieval space, thereby mitigating---but not eliminating---the risk that useful demonstrations with low Surface Lexical Similarity are excluded before delta supervision is available.

\subsubsection{Fixed Delta-Matrix Construction and Phase 2 ICL-Guided Encoder Refinement}

\begin{figure}[t]
    \centering
    \includegraphics[width=\columnwidth]{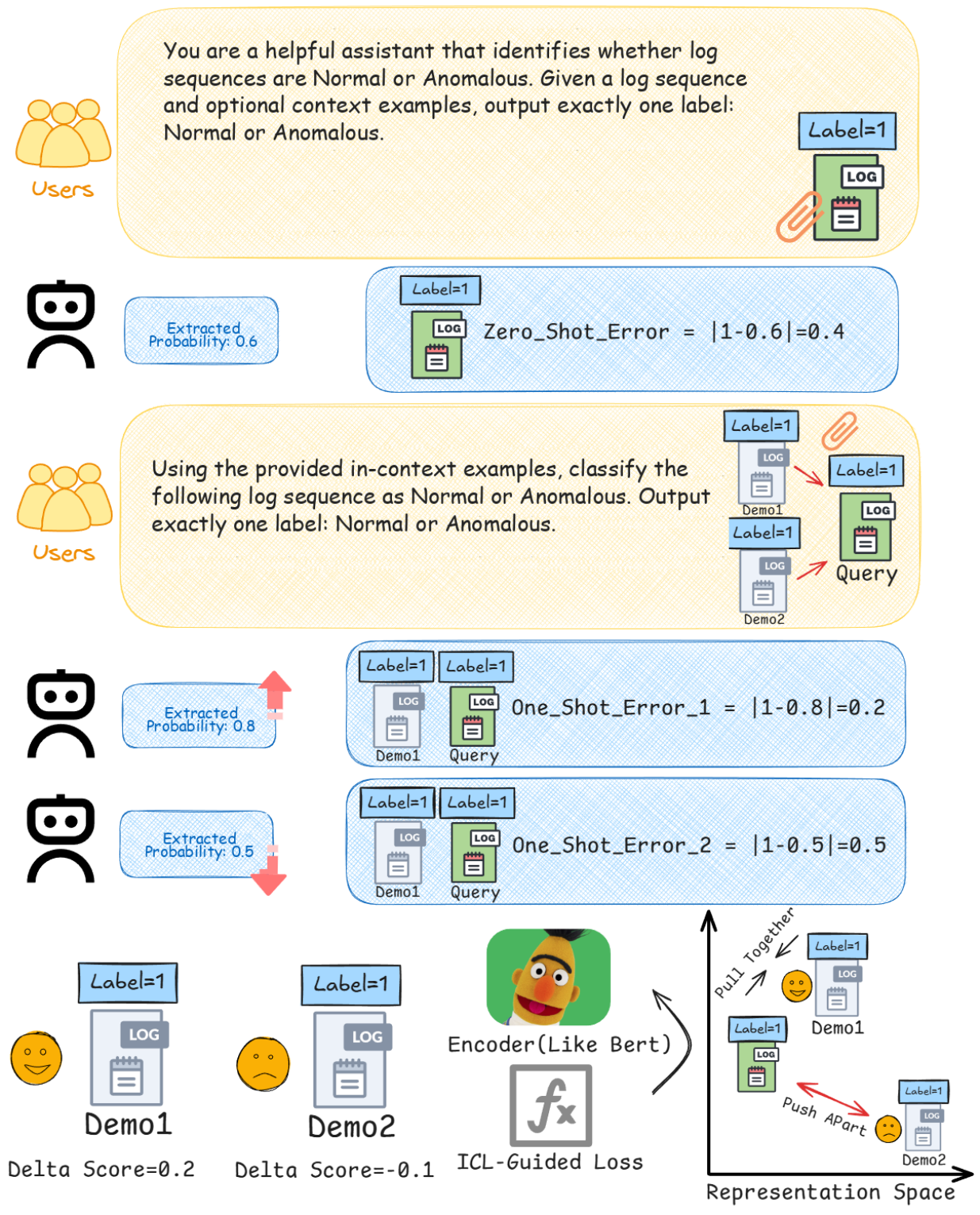}
    \caption{Illustration of fixed delta-score construction after Phase~1 and ICL-guided encoder refinement in Phase~2. For each evaluated query--demonstration pair, $\delta_{q,d}$ is computed as the reduction in the frozen LLM's prediction error from zero-shot to one-shot inference. The resulting utility signal guides the encoder to organize demonstrations in the representation space according to their observed effects on prediction.}
    \label{fig:icl_guided_refinement}
\end{figure}

After Phase~1, $f_{\boldsymbol{\theta}^{(1)}}$ is used with MMR to select candidate pairs, and the frozen LLM evaluates their demonstration utilities to construct the fixed delta matrix. As illustrated in Fig.~\ref{fig:icl_guided_refinement}, for each MMR-selected pair, the query's zero-shot prediction is compared with a separate one-shot prediction conditioned on that demonstration. Once the delta matrix is fixed, Phase~2 uses these scores to supervise encoder refinement through the second branch of Eq.~(\ref{eq:overall_training_loss}).

\paragraph{Demonstration Selection}
Evaluating all ordered query--demonstration pairs requires $O(N^2)$ LLM calls. We instead encode each training sequence as

\begin{equation}
v_i^{(1)}=f_{\boldsymbol{\theta}^{(1)}}(s_i).
\end{equation}

For each query $s_q$, Maximum Marginal Relevance (MMR) constructs a candidate set $\mathcal{D}_q$ of size $K$ by balancing relevance and diversity~\cite{carbonell1998use}. Here, $\mathcal{D}_{\mathrm{train}}=\{s_i\mid(s_i,y_i)\in\mathcal{S}_{\mathrm{train}}\}$ denotes the set of training log sequences, distinct from the labeled-pair set $\mathcal{S}_{\mathrm{train}}$. Given the current selected set $S$, let $\mathcal{A}_q(S)=\mathcal{D}_{\mathrm{train}}\setminus(S\cup\{s_q\})$. Starting from $S=\varnothing$, MMR iteratively selects

\begin{equation}
\label{eq:mmr}
\begin{aligned}
s^{*}=\operatorname*{arg\,max}_{s_j\in\mathcal{A}_q(S)}
\big\{&\lambda_{\mathrm{MMR}}
\operatorname{sim}\!\left(v_q^{(1)},v_j^{(1)}\right) \\
&-(1-\lambda_{\mathrm{MMR}})
\max_{s_k\in S}
\operatorname{sim}\!\left(v_j^{(1)},v_k^{(1)}\right)
\big\}.
\end{aligned}
\end{equation}

The diversity term is zero for $S=\varnothing$, and selection continues until $|S|=K$, at which point $\mathcal{D}_q\leftarrow S$. We use $\lambda_{\mathrm{MMR}}=0.2$ and $K=128$. Unlike pure top-$K$ similarity retrieval, MMR suppresses near-duplicate candidates and broadens the pair set considered for utility evaluation. Here, $K$ is the sparsification budget for delta construction, not the number of demonstrations used in the final prompt.

\paragraph{Computation of Delta Scores}
For each $s_d\in\mathcal{D}_q$, let $p_0(s_q)$ and $p_1(s_q\mid s_d)$ be the frozen LLM's zero-shot and one-shot anomaly probabilities. Given the ground-truth label $y_q$, we define

\begin{equation}
\label{eq:delta}
\begin{aligned}
e_0(s_q)&=|p_0(s_q)-y_q|,\\
e_1(s_q,s_d)&=|p_1(s_q\mid s_d)-y_q|,\\
\delta_{q,d}&=e_0(s_q)-e_1(s_q,s_d).
\end{aligned}
\end{equation}

A positive $\delta_{q,d}$ indicates reduced prediction error, whereas a negative value indicates interference. Thus, delta measures pairwise predictive utility, not a generated rationale or an internal reasoning state. The probabilities are obtained by applying a two-way softmax to the first-position label-token log-probabilities for \{``Normal'', ``Anomalous''\}; the deterministic decoding configuration is specified in Section~IV-A-6.

Let $\mathcal{Q}\subseteq\{1,\ldots,N\}$ be the set of unique query representatives after the implemented cache-key deduplication, and let $N_q=|\mathcal{Q}|$. A repeated training-sequence instance is mapped to the representative of its first occurrence and reuses that cached Delta row. Let $\Omega=\{(q,d):q\in\mathcal{Q},\,s_d\in\mathcal{D}_q\}$ be the newly evaluated pair set. Its scores and observation mask $\mathbf{M}\in\{0,1\}^{N_q\times N}$ are

\begin{equation}
\begin{aligned}
(\mathbf{M}_{\delta})_{qd}&=\delta_{q,d}, &&(q,d)\in\Omega,\\
\mathbf{M}_{qd}&=\mathbb{I}[(q,d)\in\Omega].
\end{aligned}
\end{equation}

Only the $|\Omega|=N_qK$ newly evaluated triples $(q,d,\delta_{q,d})$ are materialized; repeated query instances reuse their representative's cached row and trigger no additional LLM evaluation. Unobserved entries remain masked rather than being assigned $\delta=0$. The resulting $O(N_qK)$ construction fixes $\Omega$, $\mathbf{M}_{\delta}$, and $\mathbf{M}$ for Phase~2: candidate selection is not refreshed, and Delta scores are not recomputed as the encoder changes.

\paragraph{ICL-Guided Loss}
For the current encoder, let $v_q=f_{\boldsymbol{\theta}}(s_q)$, $v_d=f_{\boldsymbol{\theta}}(s_d)$, and $s_{q,d}=\operatorname{sim}(v_q,v_d)\in[-1,1]$. Positive and negative utility pairs are defined only over $\Omega$:

\begin{equation}
\label{eq:positive_negative_pairs}
\begin{aligned}
D^{+}&=\{(q,d)\in\Omega:\delta_{q,d}>0\},\\
D^{-}&=\{(q,d)\in\Omega:\delta_{q,d}<-\theta\},
\end{aligned}
\end{equation}

where $\theta\geq0$ is a dead-zone threshold; pairs with $\delta_{q,d}\in[-\theta,0]$ contribute no gradient. For numerical completeness, let $n_{+}=\max\{1,|D^{+}|\}$ and $n_{-}=\max\{1,|D^{-}|\}$. The ICL-guided loss is

\begin{equation}
\label{eq:icl_loss}
\begin{aligned}
\mathcal{L}_{\Delta}
={}&
\frac{1}{n_{+}}
\sum_{(q,d)\in D^{+}}
\delta_{q,d}
\,
\softplus\!\left(
-\frac{s_{q,d}}{\tau}
\right)\\
&+
\lambda
\frac{1}{n_{-}}
\sum_{(q,d)\in D^{-}}
w_{q,d}
\,
\softplus\!\left(
\frac{s_{q,d}}{\tau}
\right),
\end{aligned}
\end{equation}

where $w_{q,d}=\max(0,|\delta_{q,d}|-\theta)$ and $\softplus(x)=\log(1+e^x)$. If $D^{+}$ or $D^{-}$ is empty in a training batch, the corresponding empty sum and loss term are defined as zero. Minimizing $\mathcal{L}_{\Delta}$ attracts positive-utility pairs and repels sufficiently interfering pairs, with strength determined by their delta magnitudes. Only observed pairs in $\Omega$ contribute gradients. Initialized with the Phase~1 parameters $\boldsymbol{\theta}^{(1)}$, this optimization yields the final retrieval encoder $f_{\boldsymbol{\theta}^{(2)}}$ used in the test phase.

\subsection{Test Phase}

During inference, we employ a \textit{Dual-Source Retrieval Strategy} to construct optimal in-context demonstrations. For a given test log $s_t$, the final Phase~2 encoder produces \(v_t=f_{\boldsymbol{\theta}^{(2)}}(s_t)\), and each training sequence $s_j$ is represented as \(v_j=f_{\boldsymbol{\theta}^{(2)}}(s_j)\). The selection process proceeds in two stages to balance surface lexical similarity and latent semantic equivalence:

\subsubsection{Similarity-based Anchoring}
We first retrieve the \textit{top-i} nearest neighbors from the training set $\mathcal{D}_{\mathrm{train}}$ based on cosine similarity to $s_t$. These samples form the similarity-based anchor set $\mathcal{C}_{\mathrm{sim}}$:
\begin{equation}
\label{eq:anchor_sim}
\setlength{\abovedisplayskip}{3pt}
\setlength{\belowdisplayskip}{3pt}
\mathcal{C}_{\mathrm{sim}} = \mathop{top\textit{-}i}_{s_j \in \mathcal{D}_{\mathrm{train}}} \left( \frac{v_t \cdot v_j}{\|v_t\| \|v_j\|} \right)
\end{equation}

\subsubsection{Delta-guided Expansion}
We leverage the retrieved anchors in $\mathcal{C}_{\mathrm{sim}}$ to query the precomputed delta matrix. The underlying assumption is that optimal demonstrations for the anchors are likely effective for the test sample due to their high similarity. Specifically, we select the \textit{top-j} training samples that yielded the highest delta scores for the anchors in $\mathcal{C}_{\mathrm{sim}}$. We aggregate these precomputed scores and select the \textit{top-j} samples to form the expansion set $\mathcal{C}_{\delta}$:

\begin{equation}
\label{eq:delta_expansion}
\setlength{\abovedisplayskip}{3pt}
\setlength{\belowdisplayskip}{3pt}
\begin{aligned}
g(n)
&=\sum_{s_m\in\mathcal{C}_{\mathrm{sim}}}
\mathbf{M}_{m,n}\,\mathbf{1}[\delta_{m,n}>0],\delta_{m,n},\\
\mathcal{C}_{\delta}
&=\mathop{top\textit{-}j}_{\substack{s_n\in\mathcal{D}_{\mathrm{train}}\\
\sum_{s_m\in\mathcal{C}_{\mathrm{sim}}}
\mathbf{M}_{m,n}\,\mathbf{1}[\delta_{m,n}>0]>0}}
g(n).
\end{aligned}
\end{equation}

\noindent where $\delta_{m,n}$ represents the precomputed delta score of using $s_n$ as a demonstration for anchor $s_m$, $\mathbf{M}_{m,n}$ indicates whether this anchor--candidate relation was evaluated during Delta-matrix construction, and $\mathbf{1}[\delta_{m,n}>0]$ is one only for a positive delta score and zero otherwise. Because the Delta matrix is sparse, only observed positive relations contribute to $g(n)$. An unobserved or non-positive relation contributes no term, and a candidate is eligible for $\mathcal{C}_{\delta}$ only when it has at least one observed positive relation to the retrieved anchors. Finally, the LLM performs anomaly detection using the combined context $\mathcal{C}_{\mathrm{final}} = \mathcal{C}_{\mathrm{sim}} \cup \mathcal{C}_{\delta}$.

\subsubsection{In-context Prediction and Optional Diagnostic Explanation}
The demonstrations $\mathcal{C}_{\mathrm{final}}=\{s_1,s_2,\ldots,s_k\}$ and their labels $\{y_1,y_2,\ldots,y_k\}$ are concatenated with the test query $s_t$ to construct the prompt.

\begin{equation}
\mathcal{P}_t = \{(s_{1}, y_{1}), (s_{2}, y_{2}), \dots, (s_{k}, y_{k}), s_t\}.
\end{equation}

The prompt $\mathcal{P}_t$ is then provided to the LLM to estimate the anomaly probability of the test log $s_t$:
\begin{equation}
p_t = \textit{LLM}(\mathcal{P}_t),
\end{equation}
where \(p_t \in [0,1]\) denotes the probability that \(s_t\) is anomalous.  
When a diagnostic explanation is required, an optional chain-of-thought-style prompt can be applied after demonstration retrieval. This optional generation setting is not used to construct the delta matrix or to train the demonstration retriever.

\section{Evaluation}

\subsection{Experimental Design}
\subsubsection{Datasets}
We evaluate our method on four public log benchmarks: HDFS, BGL, Thunderbird, and Liberty. These datasets encompass diverse system architectures and data distributions, providing a heterogeneous operational testbed for evaluating cross-domain generalization across the selected logging systems~\cite{oliner2007supercomputers}.

\textbf{Distributed File System Logs:} Represented by HDFS, this dataset is collected from 200 Amazon EC2 nodes and contains 11,175,629 log messages \cite{xu2009detecting}.

\textbf{Supercomputing System Logs:} This category includes BGL, Thunderbird, and Liberty, which are derived from high-performance computing environments but exhibit distinct alert characteristics. BGL is dominated by hardware alerts (98.04\%).
In contrast, Thunderbird and Liberty primarily feature software-related alerts \cite{oliner2007supercomputers}.

\subsubsection{Data Preprocessing and Sampling}
We adopt distinct log grouping strategies based on the dataset structure. For HDFS, log messages are grouped by their unique \textit{block\_IDs}. This grouping produces 575,061 log sequences, of which 16,838 are anomalous, corresponding to a sequence-level anomaly ratio of 2.93\%. HDFS sequences are ordered by the first occurrence of each block ID in the structured log. For the supercomputing datasets (BGL, Thunderbird, and Liberty), we employ fixed non-overlapping windows, setting the window size to 40 for BGL and Thunderbird and 30 for Liberty. Their grouped sequences preserve the chronological order of the original log messages. The structured BGL file used in our experiments contains 4,713,493 log messages and forms 117,838 sequences, of which 24,697 are anomalous (20.96\%); at the log-line level, 824,823 messages are anomalous (17.50\%). In few-shot transfer, the first 5,000 target-domain sequences in the stated order form the target-domain training subset, and every remaining target-domain sequence forms the test set. The resulting few-shot test-set sizes are 112,838 for BGL, 495,000 for Thunderbird, and 161,667 for Liberty. In target-domain zero-shot transfer, no target-domain sequence is removed or reserved before evaluation; the complete target dataset is used as the test set. The resulting target-domain zero-shot test-set sizes are therefore 117,838 for BGL, 500,000 for Thunderbird, 166,667 for Liberty, and 575,061 for HDFS. Source-domain training subsets contain the first 50,000 or 5,000 source sequences in the same dataset-specific order.

\begin{table}[htbp]
    \centering
    \small % 保持小号字体
    \setlength{\tabcolsep}{5.5pt} % 适度收紧列间距，避免新增 BGL 双行造成轻微横向溢出
    
    % 保持标题样式为小型大写 + 无句号
    \caption{\textsc{Details of the datasets after preprocessing}}
    \label{tab:dataset_details}
    
    % 修改处：将第一个 l 改为 c，使 Dataset 列居中
    \begin{tabular}{@{} c c c c @{}}
        \toprule
        \textbf{Dataset} & \textbf{Log Messages} & \textbf{Log Seq} & \textbf{Anomaly Log Seq} \\
        \midrule
        HDFS & 11,175,629 & 575,061 & 16,838 \\
        BGL & 4,713,493 & 117,838 & 24,697 \\
        Thunderbird & 20,000,000 & 500,000 & 154,108 \\
        Liberty & 5,000,000 & 166,667 & 110,427 \\ 
        \bottomrule
    \end{tabular}
\end{table}

\subsubsection{Cross-domain Evaluation Settings}
To evaluate the generalization capability of LogICL across HDFS, BGL, Thunderbird (TB), and Liberty, we design two settings with strictly defined data access. For BGL, Thunderbird, and Liberty, sequence order follows the chronological order of the original logs after fixed non-overlapping grouping; HDFS sequence order follows the first occurrence of each block ID in the structured log. (1) In few-shot transfer, the training data combine the first 50,000 source-domain sequences with the first 5,000 target-domain sequences (e.g., Liberty--BGL), as shown in Table~\ref{tab:few_shot_results}. During inference, $k=8$ demonstrations are retrieved from this training pool to predict every remaining target-domain sequence. (2) In target-domain zero-shot transfer, no labeled or unlabeled target-domain data are accessed before evaluation. The training data instead combine the first 50,000 sequences from the primary source domain with the first 5,000 sequences from an auxiliary source domain. As shown in Table~\ref{tab:zero_shot_results} (e.g., BGL--Liberty--TB), the model retrieves $k=8$ demonstrations from these two source-domain subsets and predicts every sequence in the complete target dataset.

\subsubsection{Baselines}
To comprehensively evaluate the effectiveness of LogICL, we benchmark it against ten representative baselines across three major paradigms:

\textbf{Deep Learning-based Approaches:} We include LogRobust \cite{zhang2019robust} as a standard supervised baseline. This method transforms distinct log events into semantic vectors and employs a Long Short-Term Memory (LSTM) network to capture temporal patterns and contextual semantics, enabling stable anomaly detection across variable log data. 

\textbf{Cross-System Approach:} For cross-system scenarios, we adopt LogTransfer~\cite{chen2020logtransfer}, LogTAD~\cite{han2021unsupervised}, and MetaLog~\cite{zhang2024metalog}, which utilize transfer learning and meta-learning techniques to adapt to target domains.

\textbf{LLM-based Approaches:} This category assesses the effectiveness of LLM-based strategies, comprising both fine-tuned and parameter-frozen approaches. LogLLM first encodes log sequences into feature vectors and then performs supervised fine-tuning of the LLM on normal-only target-domain sequences for anomaly detection. LogGPT~\cite{han2023loggpt} trains a GPT-style decoder on normal-only sequences for next-event prediction. Among the frozen models, LogPrompt uses training-free prompting and relies on the LLM's intrinsic knowledge. LogRAG \cite{zhang2024lograg} is a hybrid unsupervised method combining a DeepSVDD classifier with Retrieval-Augmented Generation (RAG) technology~\cite{lewis2020retrieval}. To assess the efficacy of off-the-shelf ICL strategies in the log text domain, we adapt two standard baselines widely used in general NLP tasks~\cite{qin2024context, peng2024revisiting}. Random-ICL selects demonstrations via uniform random sampling from the candidate pool, serving as a stochastic baseline to measure the impact of demonstration content. $k$NN-ICL retrieves the top-$k$ nearest neighbors based on cosine similarity in the embedding space, representing a standard retrieval-augmented approach without domain-specific optimization.

\subsubsection{Evaluation Metrics}
We choose Precision, Recall, and F1 score as evaluation metrics, with the definitions of $\mathit{Precision} = \frac{\mathit{TP}}{\mathit{TP}+\mathit{FP}}$, $\mathit{Recall} = \frac{\mathit{TP}}{\mathit{TP}+\mathit{FN}}$, and $\mathit{F1} = \frac{2 \cdot \mathit{Precision} \cdot \mathit{Recall}}{\mathit{Precision} + \mathit{Recall}}$. Here, $\mathit{TP}$, $\mathit{FP}$, and $\mathit{FN}$ denote the number of true positives, false positives, and false negatives, respectively. Tables~II--IV report point estimates under the stated fixed data protocol; these values are interpreted as performance under that protocol rather than as evidence of statistical robustness or statistical significance.

\subsubsection{Implementation Details}
We implement our framework in PyTorch with Python 3.11, accelerated by two NVIDIA H200 GPUs (141~GB). We employ Sentence-BERT as the backbone encoder. It maps log sequences to a 384-dimensional dense vector space, while Qwen3-14B~\cite{yang2025qwen3} serves as the core LLM and is locally served via vLLM. For Delta-matrix construction, the LLM is queried with \texttt{guided\_choice} decoding (temperature $0$, \texttt{max\_tokens}$=8$, reasoning disabled), and anomaly probabilities are read from first-position label-token log-probabilities through a two-way softmax. This configuration is important for interpreting the supervision: the Delta score is computed from output label probabilities and represents demonstration utility at the prediction level, rather than a distilled chain-of-thought trace or hidden reasoning representation. Query instances are deduplicated using the implemented \texttt{item\_EventId}-based cache key before LLM evaluation. In the TLL few-shot run, 55,000 training-sequence instances yield 50,583 unique query representatives (46,240 source and 4,343 target); the remaining 4,417 repeated instances reuse the cached Delta row of their first occurrence and trigger no additional LLM call. MMR retrieves 128 candidates for each unique query representative. At test time, eight demonstrations are retrieved for each query, and exactly one Qwen3-14B call produces the final anomaly prediction. The multi-objective weights are set to $\lambda_{\mathrm{MMD}}=0.1$, $\lambda_{\mathrm{SupCon}}=1.0$, and $\lambda_{\Delta}=1.0$; the demonstration-utility loss uses $\theta=0.2$, $\lambda=0.5$, and $\tau=1.0$. Consistent with standard practices, we apply a classification threshold of 0.5 to determine anomaly labels. For all result tables, Precision, Recall, and F1 are computed from the same underlying prediction counts and independently rounded to two decimal places; consequently, recomputing F1 from the displayed Precision and Recall may occasionally produce a difference of 0.01 percentage points. \textbf{Our code is open-sourced at} \url{https://zenodo.org/records/17850648}.

\begin{table*}[htbp]
\centering
\caption{\textsc{Evaluation under Cross-Domain Few-Shot Settings (Precision, Recall, F1-Score).}}
\label{tab:few_shot_results}
\renewcommand{\arraystretch}{1.25} % 适当增加行高
\setlength{\tabcolsep}{4pt} % 微调列间距
\small
\resizebox{\textwidth}{!}{%
\begin{tabular}{c c ccc ccc ccc ccc}
\toprule
% 表头第一行
\multirow{2}{*}[-2pt]{\textbf{Method}} & \multirow{2}{*}[-2pt]{\textbf{Type}} 
& \multicolumn{3}{c}{\textbf{Liberty-BGL}}
& \multicolumn{3}{c}{\textbf{BGL-TB}}
& \multicolumn{3}{c}{\textbf{BGL-Liberty}}
& \multicolumn{3}{c}{\textbf{TB-Liberty}} \\

\cmidrule(lr){3-5} \cmidrule(lr){6-8} \cmidrule(lr){9-11} \cmidrule(lr){12-14}

% 表头第二行
& & P(\%) & R(\%) & F1(\%) 
& P(\%) & R(\%) & F1(\%) 
& P(\%) & R(\%) & F1(\%) 
& P(\%) & R(\%) & F1(\%) \\

\midrule
% --- Traditional / Supervised Methods ---
LogRobust & Supervised
& 39.01 & 87.06 & 53.88 
& 4.78 & 0.01 & 0.03
& 6.99 & 0.12 & 0.23
& 54.66 & 2.53 & 4.83 \\ 

LogTransfer & Supervised Cross-System
& 0.00 & 0.00 & 0.00
& 33.84 & 100.00 & 50.57
& 0.00 & 0.00 & 0.00
& 0.00 & 0.00 & 0.00 \\

LogTAD & Unsupervised Cross-System
& 62.60 & 58.66 & 60.57
& 60.77 & 61.10 & 60.93
& 86.49 & 66.24 & 75.02
& 84.41 & 67.50 & 75.01 \\

MetaLog & Supervised Cross-System
& 65.78 & 59.09 & 62.26
& 47.03 & 99.87 & 63.94
& 54.33 & 35.02 & 42.59
& 80.98 & 89.58 & 85.06 \\

\midrule % 分隔线：传统方法 vs LLM方法

% --- LLM-based Methods ---
LogLLM & LLM-based (fine-tune)
& 40.55 & 11.13 & 17.47
& 70.87 & 37.94 & 49.43
& 99.91 & 59.09 & 74.26
& - & - & - \\

LogGPT & LLM-based (fine-tune)
& 30.80 & 92.00 & 46.15
& 50.60 & 89.00 & 64.52
& 75.20 & 88.50 & 81.31
& - & - & - \\

LogPrompt & LLM-based (frozen)
& 39.35 & 76.61 & 52.00
& 51.12 & 70.65 & 59.32
& 66.62 & 47.30 & 55.32
& - & - & - \\

LogRAG & Unsupervised \& LLM (frozen)
& 21.10 & 98.56 & 34.76
& 44.66 & 100.00 & 61.74
& 88.40 & 100.00 & \textbf{93.84}
& - & - & - \\

Random-ICL & LLM-based (frozen)
& 21.34 & 100.00 & 35.17 
& 31.13 & 100.00 & 47.48 
& 68.31 & 100.00 & 81.17 
& 67.60 & 96.44 & 79.49 \\

$k$NN-ICL & LLM-based (frozen)
& 29.31 & 31.28 & 30.26 
& 40.13 & 37.85 & 38.95 
& 33.80 & 6.62 & 11.07 
& 90.58 & 53.77 & 67.48 \\

% --- 修改处：在这里加了 midrule，把 kNN-ICL 和 LogICL 隔开 ---
\midrule 

% --- Ours ---
\textbf{LogICL (Ours)} & \textbf{LLM-based (frozen)}
& 70.29 & 78.21 & \textbf{74.04}
& 91.16 & 87.54 & \textbf{89.32}
& 74.44 & 85.43 & 79.56
& 81.83 & 88.87 & \textbf{85.20} \\

\bottomrule
\end{tabular}%
}
\end{table*}

\begin{table*}[htbp]
\centering
\caption{\textsc{Evaluation under Target-Domain Zero-Shot Transfer (Precision, Recall, F1-Score).}}
\label{tab:zero_shot_results}
\renewcommand{\arraystretch}{1.25} % 保持与表1一致的行高
\setlength{\tabcolsep}{2pt} % 针对17列的密集表格，微调列间距以保持美观
\small
\resizebox{\textwidth}{!}{%
\begin{tabular}{c c ccc ccc ccc ccc ccc}
\toprule
% 表头：垂直居中 + 分组
\multirow{2}{*}[-2pt]{\textbf{Method}} & \multirow{2}{*}[-2pt]{\textbf{Type}} 
& \multicolumn{3}{c}{\textbf{BGL-Liberty-TB}} 
& \multicolumn{3}{c}{\textbf{Liberty-BGL-TB}} 
& \multicolumn{3}{c}{\textbf{TB-Liberty-BGL}} 
& \multicolumn{3}{c}{\textbf{BGL-TB-Liberty}} 
& \multicolumn{3}{c}{\textbf{TB-Liberty-HDFS}} \\

\cmidrule(lr){3-5} \cmidrule(lr){6-8} \cmidrule(lr){9-11} \cmidrule(lr){12-14} \cmidrule(lr){15-17}

& & P(\%) & R(\%) & F1(\%) 
& P(\%) & R(\%) & F1(\%) 
& P(\%) & R(\%) & F1(\%) 
& P(\%) & R(\%) & F1(\%) 
& P(\%) & R(\%) & F1(\%) \\

\midrule
LogRobust & Supervised 
& 57.36 & 91.79 & 70.60 
& 18.07 & 4.59 & 7.32 
& 66.48 & 31.69 & 42.92 
& 79.15 & 5.29 & 9.91 
& 2.43 & 82.50 & 4.73 \\

LogPrompt & LLM-based (frozen)
& 51.10 & 70.65 & 59.30 
& - & - & - 
& 39.49 & 76.72 & \textbf{52.15} 
& 66.54 & 47.30 & 55.29 
& 8.73 & 84.07 & 15.81 \\

MetaLog & Supervised Cross-System 
& 42.10 & 86.23 & 56.58 
& 56.70 & 73.50 & 64.01 
& 29.16 & 23.35 & 25.93 
& 90.72 & 10.02 & 18.04 
& 3.50 & 0.05 & 0.10 \\

\midrule
\textbf{LogICL (Ours)} & \textbf{LLM-based (frozen)} 
& 70.16 & 79.83 & \textbf{74.68} 
& 65.96 & 66.06 & \textbf{66.01} 
& 40.21 & 34.71 & 37.25 
& 82.39 & 69.70 & \textbf{75.52} 
& 72.87 & 22.75 & \textbf{34.67} \\
\bottomrule
\end{tabular}%
}
\end{table*}

\subsection{Evaluation under Cross-Domain Few-Shot Settings}

To comprehensively evaluate the effectiveness of our proposed framework, we compare LogICL with a wide range of baselines. Table~\ref{tab:few_shot_results} presents the comparative results in terms of Precision, Recall, and F1 score across four cross-domain scenarios.

% \paragraph{Baselines and Experimental Configurations}
\subsubsection{Baselines and Experimental Configurations}
It is important to clarify the specific experimental configurations adopted to ensure a fair and realistic comparison, particularly where they deviate from the original papers:

For LogRobust, we adjusted its original fully supervised setting (using an 8:2 split for training and testing) to align with our cross-domain few-shot setting. It was trained on 50,000 source-domain samples and adapted using 5,000 target-domain samples.

For LogTransfer and LogTAD, we employ the same source and target domain splits as LogICL. The training and testing sample quantities are kept strictly consistent across all experiments to ensure a fair comparison.

For the LLM-based baselines without an explicit cross-domain transfer mechanism, we follow their respective intra-domain protocols. LogLLM performs supervised fine-tuning using only the normal sequences contained in the 5,000-sequence target-domain training subset: 4,383 sequences when BGL is the target and 5,000 when Thunderbird or Liberty is the target. LogGPT likewise trains its next-event predictor on the normal sequences from this target-domain subset. LogRAG uses all 5,000 target-domain sequences in its unsupervised pipeline. LogPrompt is training-free and therefore does not use these 5,000 sequences for training. For these target-only or training-free baselines, the TB--Liberty and BGL--Liberty configurations are identical because both are evaluated on the same Liberty target data without using the source domain; therefore, the redundant TB--Liberty entries for LogLLM, LogGPT, LogPrompt, and LogRAG are omitted and marked with ``-''.

For Random-ICL and $k$NN-ICL, we construct a shared demonstration pool comprising 50,000 labeled source-domain samples and 5,000 labeled target-domain samples. During inference, demonstrations are retrieved from this pool via random sampling or $k$-nearest neighbor matching in the embedding space, respectively, and are provided alongside the test query to the LLM for anomaly prediction.

\subsubsection{Comparative Analysis and Discussion}
The results in Table~\ref{tab:few_shot_results} provide key insights into the performance of baselines under cross-domain few-shot log anomaly detection.
Traditional supervised and transfer learning methods show limited generalization. LogRobust, as a supervised model, requires ample labeled data for training and exhibits sharp declines in cross-domain scenarios (e.g., F1 of 0.03\% on BGL-TB), due to inadequate adaptation with sparse target supervision. LogTransfer, a supervised cross-system approach, is highly sensitive to data distributions; in our chronological split with only 5,000 target samples (below the 300 anomalies needed for effective transfer, as reported in the original work), it often defaults to classifying all logs as normal, yielding 0.00\% F1 in three out of four transfers (Liberty-BGL, BGL-Liberty, TB-Liberty). LogTAD and MetaLog fare better in some cases but remain inconsistent, as their reliance on pre-trained word embeddings for feature extraction captures only surface lexical similarity, overlooking latent semantic equivalence amid structural and terminological differences, which degrades accuracy under limited target data (e.g., MetaLog's 42.59\% F1 on BGL-Liberty).
LLM-based methods demonstrate greater potential but encounter specific limitations. LogLLM, which involves fine-tuning, needs extensive data to harness its large parameters for cross-domain generalization, leading to low scores with scarce target labels (e.g., 17.47\% F1 on Liberty-BGL). LogGPT, which trains a GPT-style decoder on normal-only sequences, achieves high recall (88.5\%--92.0\%) because its novelty-based detector flags any surprising next event; however, its precision is tightly coupled to anomaly density, collapsing to 30.80\% on the anomaly-light Liberty-BGL test (F1 of 46.15\%) while rising on anomaly-heavy sets (e.g., 81.31\% F1 on BGL-Liberty). Without a cross-domain alignment mechanism, legitimate-but-unseen normal patterns are misclassified as anomalies under our chronological split. Across the same three transfer settings for which LogGPT reports results, its average F1 is 63.99\%, whereas LogICL achieves 80.97\%, a difference of 16.98 percentage points. LogPrompt, using frozen LLMs, achieves moderate results but struggles without optimized prompts and demonstrations to activate the model's prior knowledge for precise log anomaly detection (e.g., 55.32\% F1 on BGL-Liberty). LogRAG exhibits substantial variation across the evaluated settings, obtaining 34.76\% F1 on Liberty-BGL and 93.84\% on BGL-Liberty. The ICL baselines, Random-ICL and $k$NN-ICL, generally underperform because random or embedding-based selection often yields irrelevant or interfering demonstrations, failing to improve downstream anomaly prediction (e.g., $k$NN-ICL's 11.07\% F1 on BGL-Liberty).

\textit{3) Effectiveness of LogICL:} Among the three frozen-LLM ICL retrieval strategies, LogICL achieves the highest F1 in three of the four transfers: $74.04\%$ on Liberty--BGL, $89.32\%$ on BGL--TB, and $85.20\%$ on TB--Liberty. On BGL--Liberty, LogICL obtains $79.56\%$ F1, below LogRAG's $93.84\%$, LogGPT's $81.31\%$, and Random-ICL's $81.17\%$; Random-ICL is $1.61$ percentage points higher than LogICL. Among the explicitly reported non-redundant entries, LogICL has the highest F1 on Liberty--BGL, BGL--TB, and TB--Liberty, whereas LogRAG has the highest F1 on BGL--Liberty. Random-ICL, $k$NN-ICL, and LogICL follow the same frozen-LLM ICL paradigm but differ in their retrieval policies. Their reported point estimates suggest that adding retrieved demonstrations alone is insufficient: learning from pairwise demonstration utility can help select cross-domain examples for downstream prediction, although the benefit is setting-dependent.

\subsection{Evaluation under Target-Domain Zero-Shot Transfer Settings}

In this section, target-domain zero-shot transfer means that no labeled or unlabeled target-domain data are accessed before evaluation; source-domain training data remain available. At evaluation time, every sequence in the complete target dataset is used as test data. Table~\ref{tab:zero_shot_results} reports LogICL together with the supervised baselines LogRobust and MetaLog and the training-free LLM inference baseline LogPrompt under this access boundary.

% \paragraph{Baselines and Experimental Configurations}
\subsubsection{Baselines and Experimental Configurations}
The supervised baselines, LogRobust and MetaLog, are trained on a composite dataset comprising 50,000 sequences from the primary source and 5,000 from the auxiliary source and are then evaluated on the unseen target domain without target-support adaptation. LogPrompt keeps the LLM parameters frozen and uses only the current target test query at inference time; it does not use a target-domain training or support subset. Because its input is independent of the source-domain configuration, redundant LogPrompt entries are marked with ``-''.

\subsubsection{Comparative Analysis and Discussion}
The results in Table~\ref{tab:zero_shot_results} highlight the generalization limitations of the compared supervised and meta-learning approaches. LogRobust uses static semantic embeddings and a supervised classifier trained only on the two source subsets; its F1 drops to 4.73\% on TB--Liberty--HDFS. MetaLog is likewise trained on the 50,000 primary-source and 5,000 auxiliary-source sequences and is evaluated without the target-support adaptation used by its standard meta-learning procedure. Its low F1 in this setting (e.g., 0.10\%) therefore reflects limited transfer from the available source domains when no target-domain support set is provided.
Among the prompt-based results, LogPrompt surpasses LogICL on TB--Liberty--BGL (52.15\% vs. 37.25\% F1), a difference of 14.90 percentage points, but obtains 15.81\% F1 when HDFS is the unseen target. These point estimates show that its performance varies substantially across the evaluated transfers; without targeted demonstration retrieval, it can struggle with fine-grained anomaly patterns in structurally complex datasets such as HDFS.
In contrast, LogICL uses LLM-derived demonstration utility to train the encoder and retrieves source-domain demonstrations for frozen-LLM prediction, enabling cross-domain transfer without target-domain supervision.
It is worth noting that the TB-Liberty-HDFS transfer represents the most challenging cross-domain scenario in our evaluation, bridging logs from a supercomputing environment to a distributed storage system. Compared to the first four transfers (all within supercomputing systems), the substantial differences in log structure, terminology, and failure modes reduce the amount of transferable knowledge from the source domain. This increased domain gap explains the overall lower performance on this transfer across all methods, underscoring the difficulty of generalization in truly heterogeneous settings.

\subsubsection{Effectiveness of LogICL}
LogICL outperforms the compared baselines in four of the five target-domain zero-shot transfer settings. In the challenging TB--Liberty--HDFS setting, it achieves an F1 score of $34.67\%$, more than twice the $15.81\%$ obtained by LogPrompt. Without pre-evaluation access to target-domain data, LogICL retrieves demonstrations only from the two source-domain subsets. Compared with static prompting, the reported results indicate that source-domain demonstrations selected through LogICL's learned retrieval mechanism can provide useful contextual evidence for frozen-LLM prediction. Random-ICL and $k$NN-ICL were not evaluated in this setting, so Table~\ref{tab:zero_shot_results} does not support a target-domain zero-shot comparison with random or similarity-only retrieval.

\subsection{Parameter Sensitivity Analysis}
\begin{figure*}[htbp]
    \centering
    \label{fig:parameter_sensitivity}
    
    % --- 第一行 ---
    % 左上图
    \subfloat[BGL--Thunderbird]{
        \includegraphics[width=0.4\textwidth]{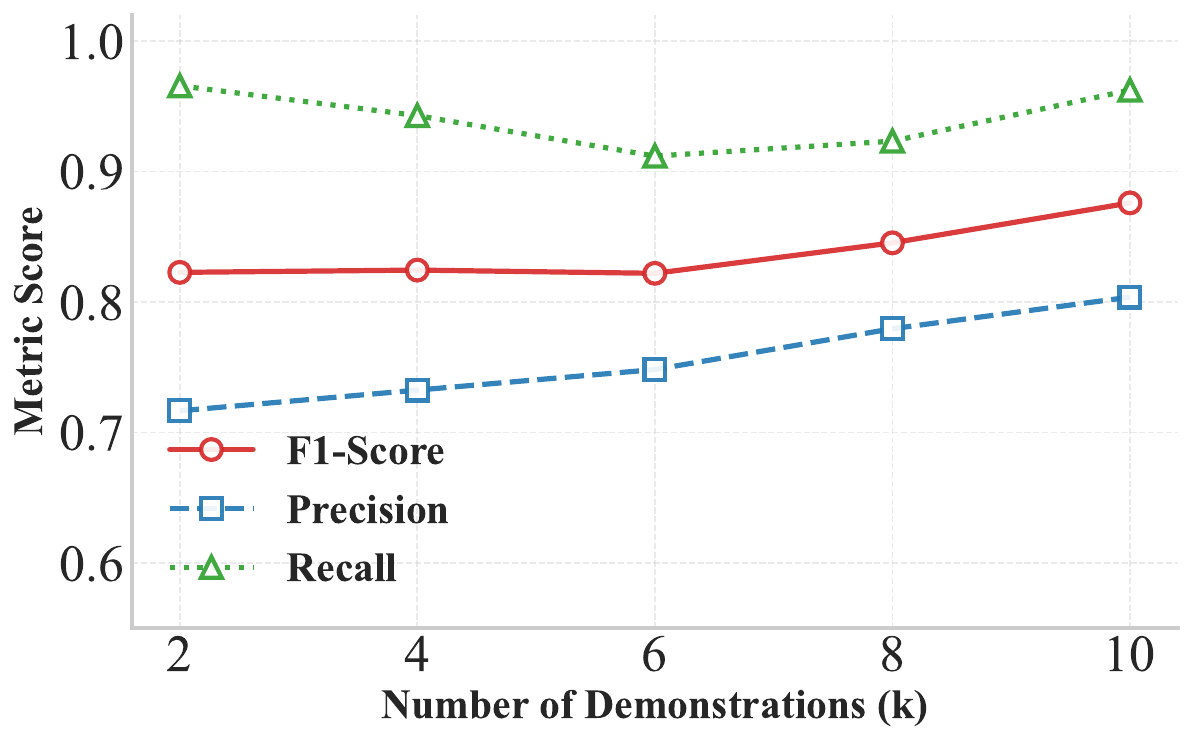}
        \label{fig:btt}
    }
    \hfill % 撑开间距
    % 右上图
    \subfloat[Thunderbird--Liberty]{
        \includegraphics[width=0.4\textwidth]{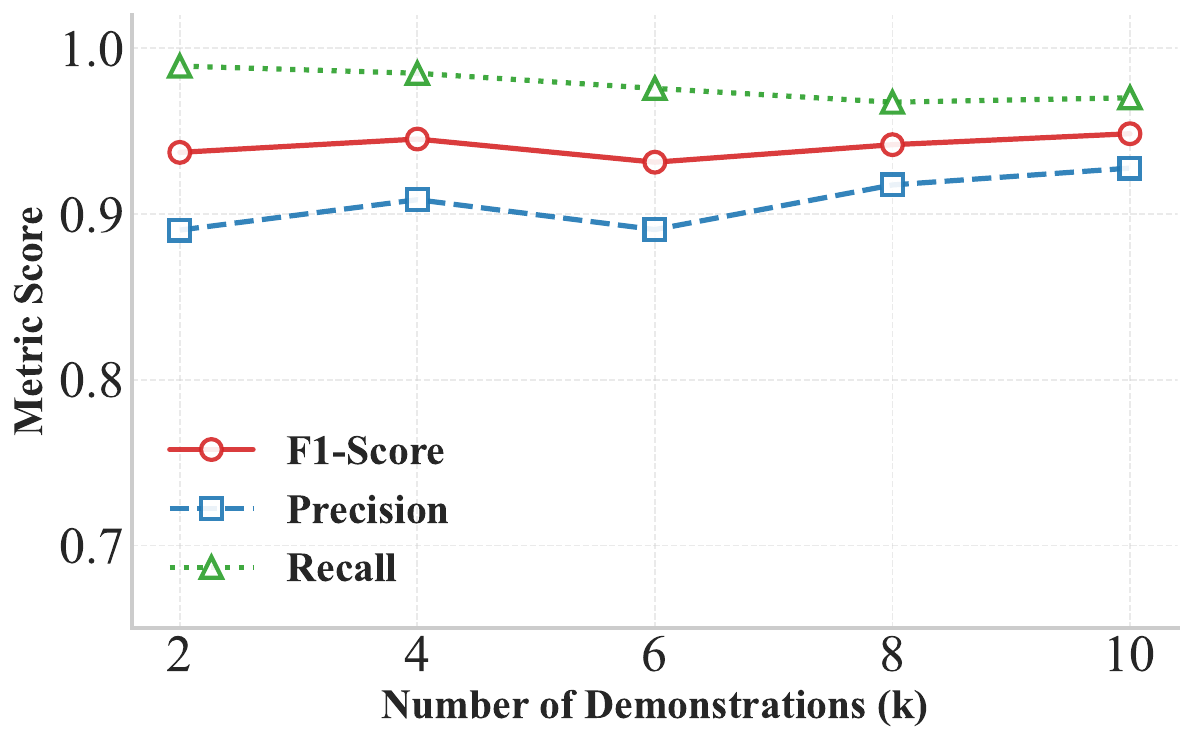}
        \label{fig:tll}
    }
    
    \vspace{-1em} % 增加行间距
    
    % --- 第二行 ---
    % 左下图
    \subfloat[BGL--Liberty]{
        \includegraphics[width=0.4\textwidth]{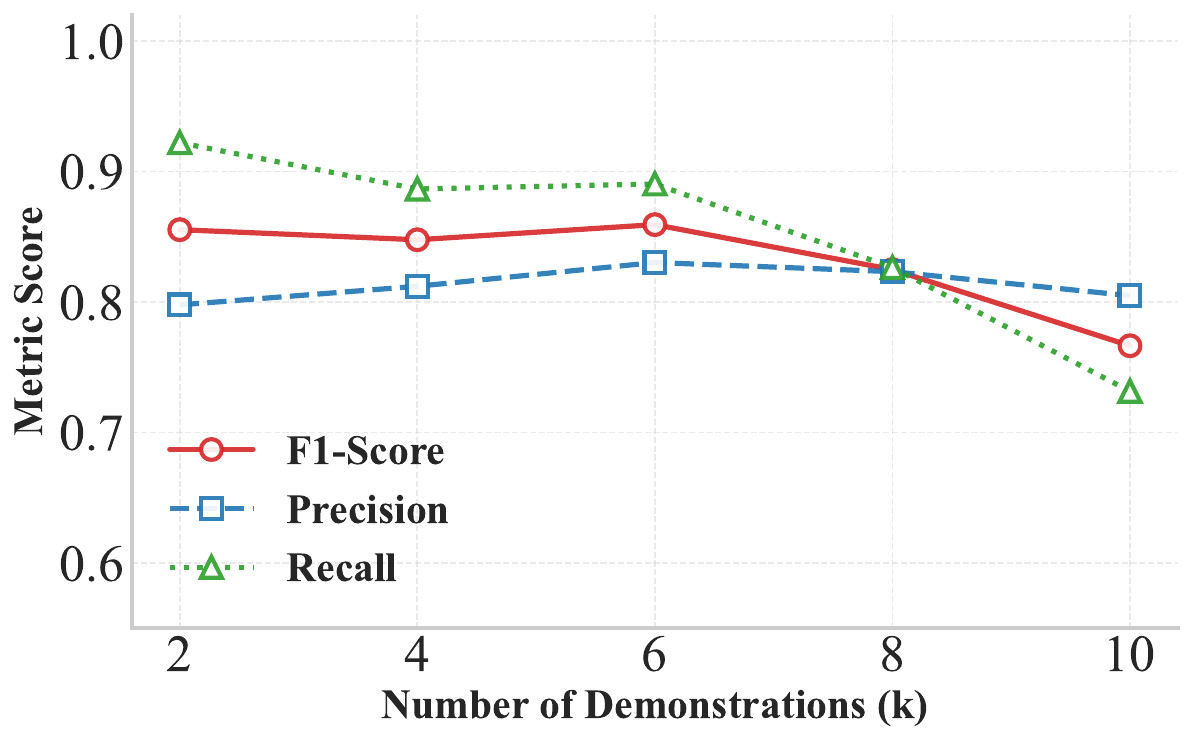}
        \label{fig:bll}
    }
    \hfill
    % 右下图
    \subfloat[Liberty--BGL]{
        \includegraphics[width=0.4\textwidth]{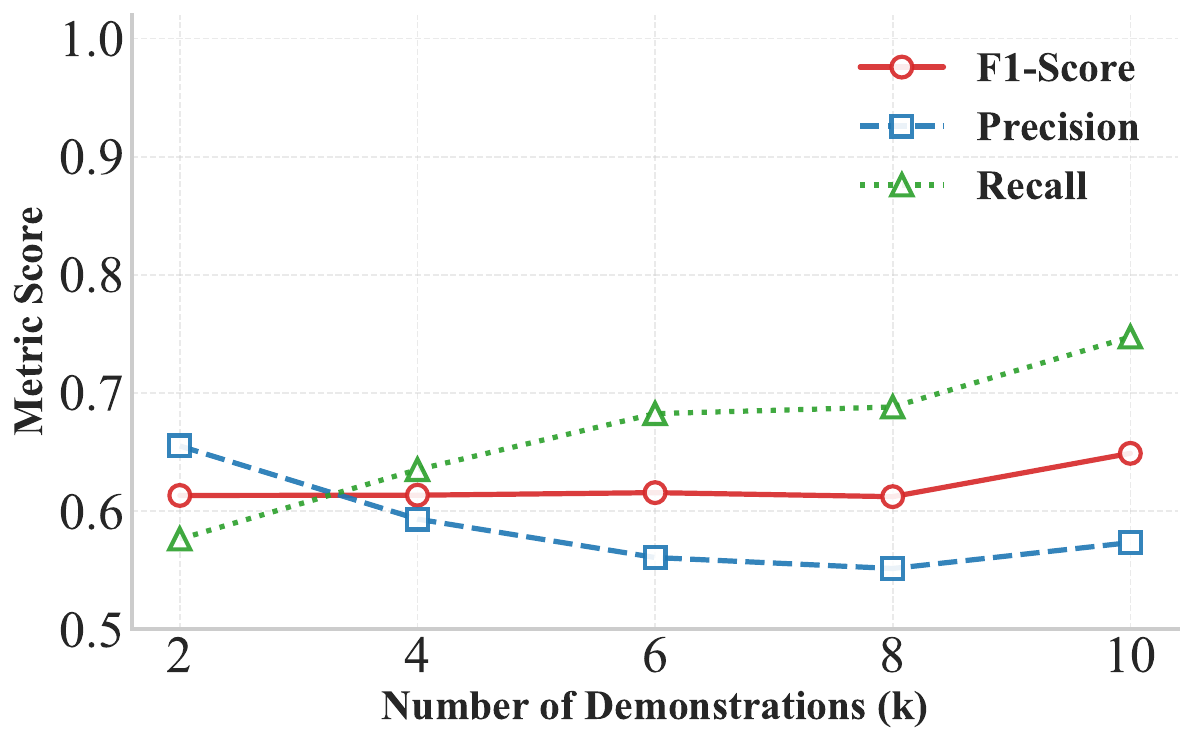}
        \label{fig:lbb}
    }
    
    \caption{Sensitivity of LogICL to the number of demonstrations across four cross-domain transfer settings.}
    \label{fig:parameter_sensitivity_all}
\end{figure*}

In this section, we investigate the sensitivity of LogICL to the number of demonstrations ($k$). We construct a randomized test set by shuffling the target-domain data and sampling 5,000 instances. The number of retrieved demonstrations $k$ is varied within the range $\{2, 4, 6, 8, 10\}$, and the classification threshold is fixed at $0.5$. Figure~\ref{fig:parameter_sensitivity_all} illustrates the trends of Precision, Recall, and F1 score across four transfer scenarios.

As observed in Figure~\ref{fig:parameter_sensitivity_all}, the effect of $k$ is setting-dependent. For Liberty--BGL (Figure~\ref{fig:parameter_sensitivity_all}(d)), the F1 score remains broadly stable for $k=2$--$8$ and increases at $k=10$. This pattern indicates that additional demonstrations can provide useful context in some settings, but it does not imply a monotonic improvement as $k$ increases.

Indeed, Recall decreases in the BGL--Liberty setting when $k>6$, as shown in Figure~\ref{fig:parameter_sensitivity_all}(c). Increasing the number of demonstrations therefore does not always improve prediction; irrelevant or redundant examples can dilute task-relevant context and degrade the frozen LLM's prediction. The observed variation is setting-dependent across the evaluated values of $k$.

%------------------------------------------------双栏图--------------------------------------------------
%---------------------------------------------------------------------------------

\begin{table*}[t]
    \centering
    \small
    \setlength{\tabcolsep}{4.5pt}
    \caption{\textsc{Component-Wise Ablation Study (F1 Score, \%).}}
    \label{tab:ablation_loss}
    \begin{tabular}{@{}l c c c c c c c@{}}
    \toprule
    \multirow{2}{*}{\textbf{Transfer Task}}
    & \multicolumn{4}{c}{\textbf{Encoder Training}}
    & \multicolumn{2}{c}{\textbf{Retrieval Pipeline}}
    & \multirow{2}{*}{\textbf{LogICL}} \\
    \cmidrule(lr){2-5}\cmidrule(lr){6-7}
    & \textbf{w/o Opt} & \textbf{w/o $\mathcal{L}_{\Delta}$}
    & \textbf{w/o MMD} & \textbf{w/o SupCon}
    & \textbf{w/o $\mathcal{C}_{\delta}$} & \textbf{w/o MMR} & \\
    \midrule
    Liberty--BGL         & 23.73 & 38.11 & 64.82 & 33.27 & 54.18 & 62.73 & \textbf{74.04} \\
    BGL--Thunderbird     & 39.58 & 79.56 & 83.47 & 57.43 & 75.82 & 82.38 & \textbf{89.32} \\
    BGL--Liberty         & 59.35 & 73.49 & 74.63 & 62.58 & 68.72 & 73.84 & \textbf{79.56} \\
    Thunderbird--Liberty & 71.95 & 83.65 & 81.57 & 72.16 & 78.72 & 81.47 & \textbf{85.20} \\
    \bottomrule
    \end{tabular}
\end{table*}

\subsection{Ablation Study}
We evaluate the components of the two-phase training and retrieval pipeline using the variants in Table~\ref{tab:ablation_loss}. w/o Opt uses the off-the-shelf encoder without task-specific optimization, whereas w/o $\mathcal{L}_{\Delta}$ retains the Phase~1 MMD and SupCon objectives but omits Phase~2 encoder refinement. For w/o MMD and w/o SupCon, the indicated Phase~1 objective is removed, after which the MMR candidate sets and delta matrix are reconstructed and Phase~2 is retrained. w/o $\mathcal{C}_{\delta}$ keeps the trained encoder fixed but replaces dual-source retrieval with similarity-only top-8 retrieval. Finally, w/o MMR replaces MMR with cosine-similarity top-$K$ during delta-matrix construction, using the same $K=128$ and LLM-evaluation budget, and then reconstructs the matrix and retrains Phase~2. All other settings remain unchanged.

The three configurations exhibit a consistent stepwise pattern across all four transfer tasks: w/o Opt $<$ w/o $\mathcal{L}_{\Delta}$ $<$ LogICL. Average F1 increases from 48.65\% with the off-the-shelf encoder (w/o Opt) to 68.70\% when Phase~1 adaptation is retained (w/o $\mathcal{L}_{\Delta}$), a gain of 20.05 percentage points, and then to 82.03\% with the complete two-phase model, a further gain of 13.33 points. Overall, the complete optimization pipeline improves average F1 by 33.38 points over w/o Opt. This consistent progression demonstrates that the domain-aligned and class-discriminative representation established in Phase~1 provides a substantially stronger basis for retrieval, while Phase~2 utility-guided refinement delivers additional gains on top of that representation.

Removing MMD or SupCon produces an end-to-end average decrease of 5.91 and 25.67 F1 points, respectively. These variants include the downstream changes induced in candidate selection, delta-matrix construction, and Phase~2 refinement. The results are consistent with their respective roles: MMD improves cross-domain distribution alignment, whereas SupCon preserves class-discriminative structure during alignment. The larger decrease without SupCon indicates that class separation is particularly important for reliable cross-domain demonstration retrieval.

At inference, replacing the dual-source strategy with similarity-only retrieval (w/o $\mathcal{C}_{\delta}$) reduces F1 by 12.67 points on average, including a 19.86-point decrease on Liberty--BGL. This result shows that delta-guided expansion contributes useful demonstrations beyond those selected solely by embedding similarity. Replacing MMR with cosine top-$K$ during delta construction decreases F1 by 6.93 points on average. Since both variants use $K=128$ and the same LLM-evaluation budget, the difference supports the value of suppressing redundant candidates and broadening the observed query--demonstration pairs. Overall, the ablations show that the two Phase~1 objectives, Phase~2 utility supervision, and the two retrieval-stage mechanisms make complementary contributions to LogICL.

%----------------------------------------------------------------------------------

% 使用 figure* 环境使图片横跨两栏
\begin{figure*}[htbp]
    \centering
    
    % --- 第一行 (Row 1) ---
    \begin{minipage}[t]{0.49\textwidth} % 左图，占据页面宽度约49%
        \centering
        % 使用 width=\linewidth 确保图片填满 minipage 的宽度
        \includegraphics[width=\linewidth]{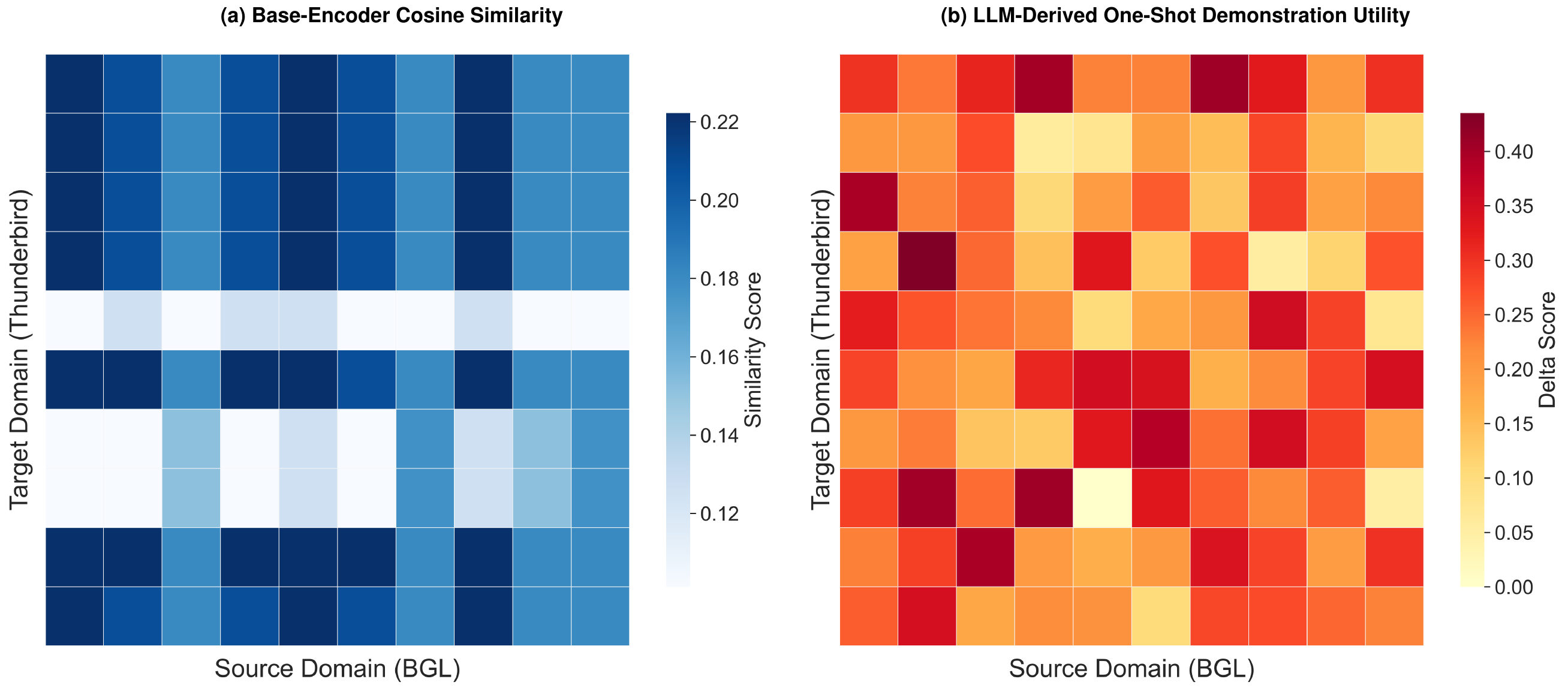} 
        \vspace{2pt}
        \centerline{\footnotesize (a) BTT Transfer} % 建议加上子图标签
    \end{minipage}
    \hfill % 填充中间的空白空间
    \begin{minipage}[t]{0.49\textwidth} % 右图，占据页面宽度约49%
        \centering
        \includegraphics[width=\linewidth]{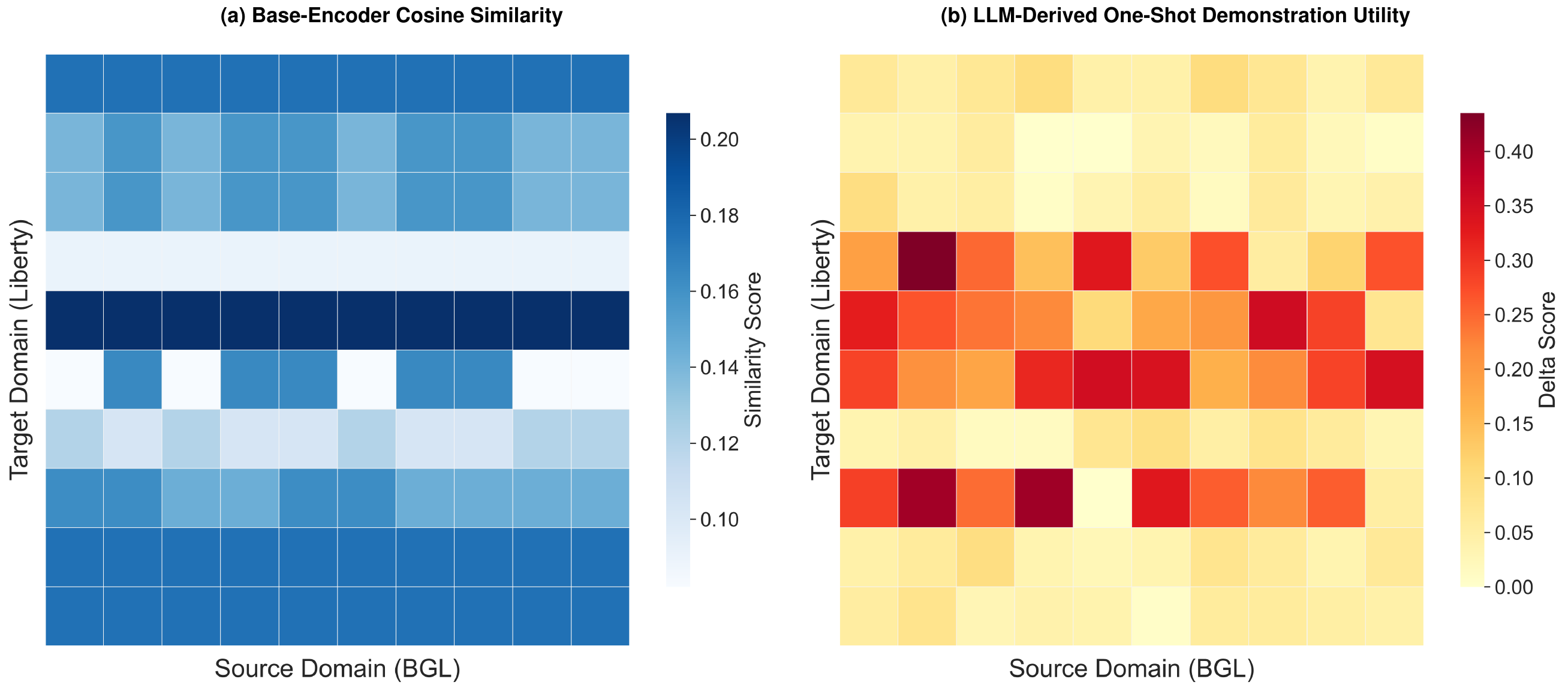}
        \vspace{2pt}
        \centerline{\footnotesize (b) BLL Transfer} % 建议加上子图标签
    \end{minipage}
    
    % 在两行之间增加一些垂直间距，让图片不至于挤在一起
    \vspace{4mm}
    
    % --- 第二行 (Row 2) ---
    \begin{minipage}[t]{0.49\textwidth} % 左图
        \centering
        \includegraphics[width=\linewidth]{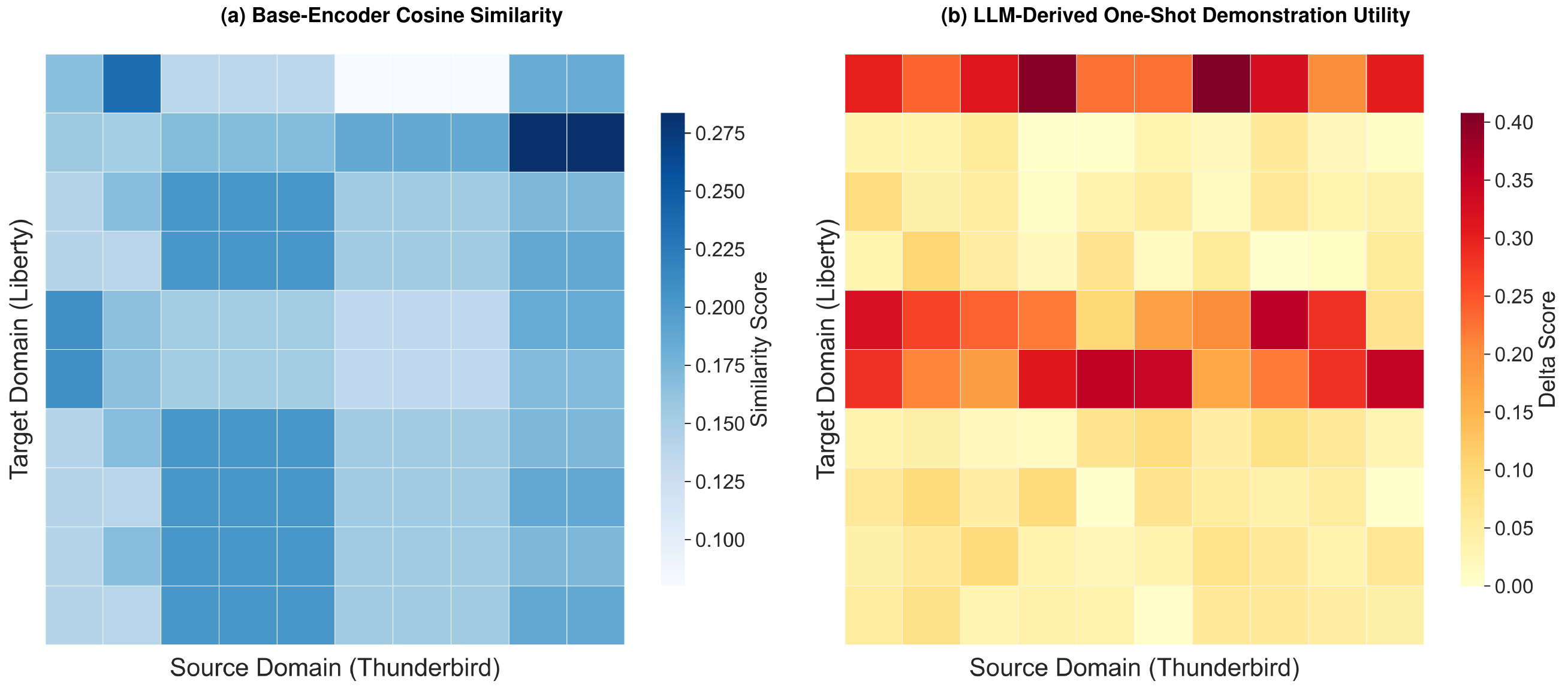}
        \vspace{2pt}
        \centerline{\footnotesize (c) TLL Transfer} % 建议加上子图标签
    \end{minipage}
    \hfill
    \begin{minipage}[t]{0.49\textwidth} % 右图
        \centering
        \includegraphics[width=\linewidth]{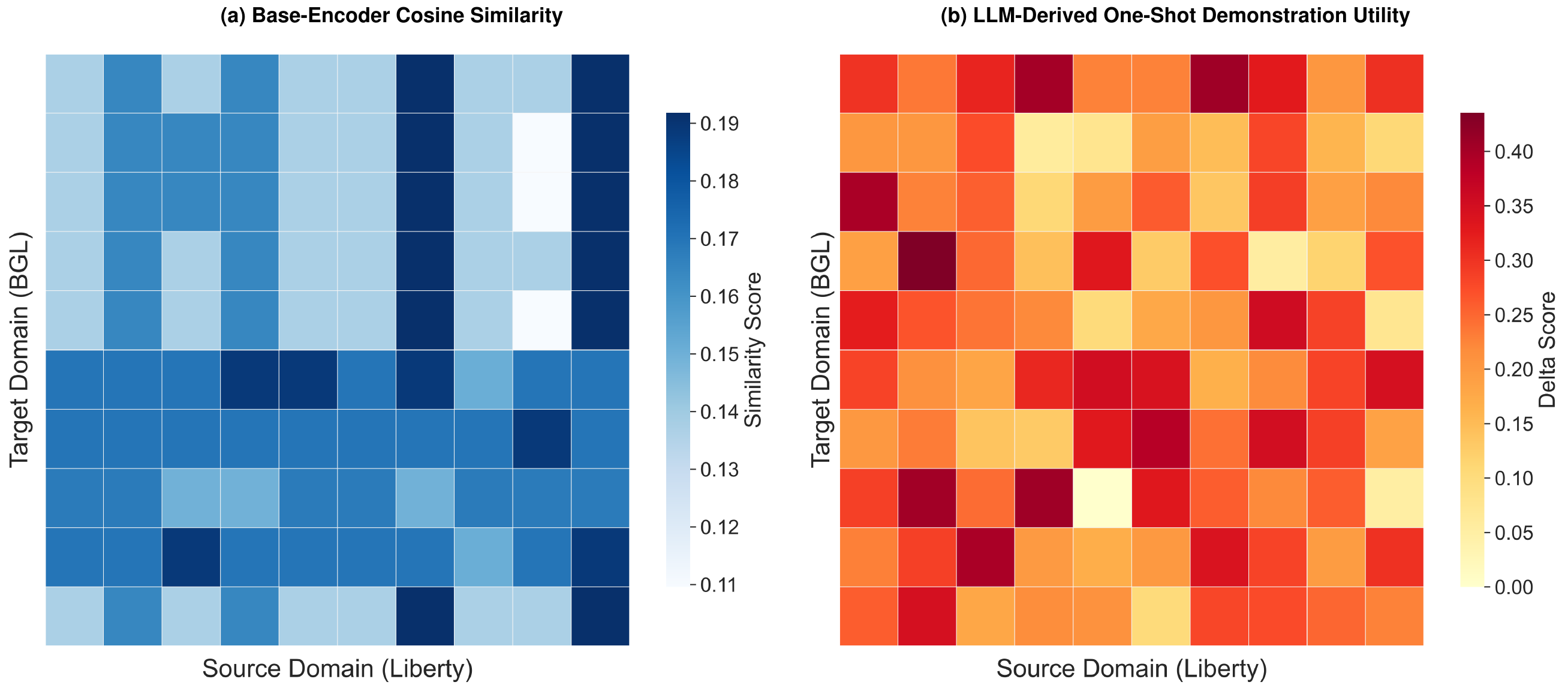}
        \vspace{2pt}
        \centerline{\footnotesize (d) LBB Transfer} % 建议加上子图标签
    \end{minipage}
    
    % 统一的图标题和标签
    \caption{Comparison of Sentence-BERT-based Surface Lexical Similarity (blue), computed as the cosine similarity between the base Sentence-BERT representations of raw log sequences, and LLM-derived one-shot demonstration utility $\delta$ (red) for evaluated cross-domain query--demonstration pairs. Darker red cells indicate larger reductions in the frozen LLM's prediction error after adding the corresponding demonstrations. The figure characterizes the utility signal used by LogICL; it does not visualize the final trained encoder.}
    \label{fig:Visualization of Alignment Mechanisms} % 建议更改标签以避免与之前的 \label{fig} 重复
\end{figure*}

\subsection{Analysis of Cross-Domain Demonstration Utility}

Figure~\ref{fig:Visualization of Alignment Mechanisms} compares two quantities for evaluated cross-domain query--demonstration pairs in four transfer settings. The blue matrices report Surface Lexical Similarity through the cosine similarity between the base Sentence-BERT representations of the raw query and demonstration sequences, whereas the red matrices report LLM-derived one-shot demonstration utility. Each red cell contains $\delta_{q,d}=e_0(s_q)-e_1(s_q,s_d)$, the reduction in the frozen LLM's prediction error after source-domain sequence $s_d$ is added as a demonstration for query $s_q$. Hence, $\delta_{q,d}>0$ indicates that the demonstration improves the corresponding prediction.

\textit{1) Surface Lexical Similarity Patterns:}
The low values in the blue matrices indicate low Surface Lexical Similarity for many visualized cross-domain pairs. The observed pattern exposes a limitation of similarity-only selection: candidates with low Surface Lexical Similarity may still improve the downstream LLM prediction.

\textit{2) Demonstration Utility Beyond Surface Lexical Similarity:}
Darker red cells indicate larger positive delta scores and therefore greater reductions in prediction error. For the visualized pairs, relatively high positive delta scores sometimes coincide with low Surface Lexical Similarity. Thus, low Surface Lexical Similarity does not necessarily imply low predictive utility for in-context anomaly detection. This observation motivates using delta as task-specific supervision rather than selecting demonstrations solely by Surface Lexical Similarity.

Importantly, Figure~\ref{fig:Visualization of Alignment Mechanisms} characterizes the utility signal collected during delta-matrix construction, which provides outcome-level supervision for subsequent delta-guided refinement. The recurring high-utility pairs with low Surface Lexical Similarity show that this utility signal captures cross-domain correspondences that are not fully reflected by Sentence-BERT cosine similarity. Together with the shared cross-domain operational patterns illustrated in Table~\ref{tab:case_study}, these results support our interpretation that LogICL can identify and exploit latent semantic equivalence beyond Surface Lexical Similarity, thereby motivating utility-aware supervision and retrieval.
%----------------------------------------------------------------------------------

\begin{table*}[t]
    \centering
    \small
    % 【修改点2】将行高从 1.5 改为 1.2，使整体更紧凑
    \renewcommand{\arraystretch}{1.2} 
    
    % 保持垂直居中设置
    \renewcommand{\tabularxcolumn}[1]{m{#1}}
    
    % 标题建议加上 \textsc 保持顶会风格的小型大写
    \caption{\textsc{Qualitative Analysis of Cross-Domain Demonstration Utility Beyond Surface Lexical Similarity.}} 
    \label{tab:case_study}
    
    % 【关键修改点】
    % 第3列原为 m{3cm}，改为 >{\centering\arraybackslash}m{3cm} 以实现居中
    \begin{tabularx}{\textwidth}{@{} 
        >{\centering\arraybackslash}m{2.5cm}   % Role (居中)
        >{\raggedright\arraybackslash}X        % Log Sequence (左对齐)
        >{\centering\arraybackslash}m{3cm}     % Surface Semantics (居中 - 已修改)
        m{4.5cm}                               % Shared Observed Pattern (左对齐，因为包含列表，左对齐通常更整齐)
    @{}}
    \toprule
    \textbf{Role} & \textbf{Log Sequence Context} & \textbf{Surface Semantics} & \textbf{Shared Observed Pattern}\newline\textbf{(Manual Interpretation)} \\
    \midrule
    
    % --- Row 1: Target Input ---
    \textbf{Target Domain} (BGL) & 
    generating core.2516 ;-; generating core.2517 ;-; generating core.2380 ;-; generating core.2382 ;-; generating core.2509 ... & 
    \textbf{Process Crash} \newline \textcolor{gray}{(Memory Context)} & 
    
    % 右侧跨行
    \multirow{2}{=}{\footnotesize%
        \textbf{Shared Operational Pattern:}\par
        1. \textbf{Availability Disruption}\newline
        {\color{gray}\scriptsize (Process Crash / Resource Allocation Failure)}\par
        2. \textbf{Burst-Like Repetition}\newline
        {\color{gray}\scriptsize (Repeated Failure-Related Events)}
    } \\
    
    % 分隔线
    \cmidrule(r){1-3}
    
    % --- Row 2: Retrieved Demo ---
    \textbf{Retrieved Demo} (Thunderbird) & 
    dhcpd: DHCPDISCOVER from ... network 172.30.0/16: \textcolor{red}{no free leases} ;-; dhcpd: DHCPDISCOVER from ... network 172.31.0/16: \textcolor{red}{no free leases} ... & 
    \textbf{Alloc. Failure} \newline \textcolor{gray}{(Network Context)} & 
    \\ 
    
    \bottomrule
    \end{tabularx}
\end{table*}

\subsection{Case Study}
To examine delta-guided retrieval qualitatively, we analyze a representative pair with low Surface Lexical Similarity (Sentence-BERT cosine similarity $<0.1$) and high positive utility ($\delta>0.8$). As shown in Table~\ref{tab:case_study}, the BGL query contains repeated \texttt{generating core} events, whereas the retrieved Thunderbird demonstration contains repeated \texttt{dhcpd: no free leases} messages. Adding this demonstration reduces the frozen LLM's prediction error despite the low Surface Lexical Similarity. Manual inspection identifies two shared cross-domain operational patterns: availability disruption and burst-like repetition. Although these patterns are manifested as a process crash in BGL and a resource-allocation failure in Thunderbird, both sequences contain repeated abnormal events associated with reduced system or service availability. These shared higher-level failure characteristics provide a plausible post hoc explanation for the demonstration's positive utility despite the pair's low Surface Lexical Similarity. The example therefore illustrates that the trained retriever can prioritize a useful cross-domain demonstration by exploiting related failure semantics beyond Surface Lexical Similarity; this interpretation describes the observed cross-domain relation rather than the LLM's internal reasoning process.

\subsection{Threats to Validity}
We acknowledge four primary factors that may impact the validity and generalization of our findings.

\textbf{Sensitivity to Prompt Design.} The inference accuracy of ICL is contingent upon the quality of instruction templates. As instructional ambiguity can act as a confounding variable, suboptimal prompt engineering may mask the performance gains attributed to our demonstration retrieval mechanism.

\textit{Gap between Frozen LLMs and Domain Expertise.} General-purpose LLMs do not possess complete AIOps expertise. Because LogICL does not update the LLM parameters, prediction and explanation quality remain bounded by the model's pretrained knowledge. More reliable deployment in complex diagnostic settings may require explicit domain adaptation or expert-feedback mechanisms.

\textbf{Impact of Model Scale.} Foundational research~\cite{brown2020language} characterizes the capacity for in-context learning in few-shot tasks as an emergent ability that scales significantly with model size. The 14B model used here was selected to balance accuracy and computational cost and may not attain the best ICL performance available from larger foundation models. Consequently, the reported results should not be interpreted as an upper bound on the retrieval framework's performance across LLM scales.

\textbf{Dependence on the BGL Expert-Validated Labels.} Our BGL evaluation uses the complete fixed expert-validated template-label mapping published in prior work~\cite{xu2025rationanomaly}. The same mapping is applied before data splitting and is used for LogICL and every baseline. The evaluation nevertheless depends on the validity of these external expert annotations and on the assumption that occurrences of the same mapped log template share the same operational label.

\section{Discussion}
\textbf{BGL Evaluation Labels.} We use the fixed expert-validated template-label mapping published in prior work~\cite{xu2025rationanomaly} as the BGL evaluation labels. The mapping is fixed independently of LogICL and baseline predictions, and we do not add, remove, or modify its entries according to model outputs. It is applied consistently to every method before the dataset is split.

\section{Conclusion}
In this paper, we presented LogICL, a demonstration-retrieval framework for cross-domain log anomaly detection with a frozen LLM. LogICL constructs a sparse Delta matrix from the signed change in prediction error between zero-shot and one-shot inference and uses this outcome-level signal to train a lightweight log encoder. Training is sequential: Phase~1 optimizes domain alignment and supervised contrastive learning, and Phase~2 starts from the Phase~1 encoder and refines it using the fixed LLM-derived Delta utility signal. At test time, similarity-based anchors and precomputed Delta relations are combined to retrieve demonstrations without candidate-wise LLM evaluation. Among the three frozen-LLM ICL retrieval strategies, LogICL has the highest reported F1 in three of the four few-shot transfers, while Random-ICL is $1.61$ percentage points higher on BGL--Liberty ($81.17\%$ vs. $79.56\%$). LogICL also has the highest reported F1 among the compared methods in four of the five target-domain zero-shot transfers. These fixed-split point estimates support demonstration retrieval as a useful means of transferring contextual evidence across heterogeneous logging systems; they are not presented as evidence of statistical robustness or statistical significance.

% \section*{Acknowledgments}

\bibliographystyle{IEEEtran}
\bibliography{references}{
}

\end{document}